\documentclass[11pt,preprint]{aastex}
\usepackage{graphicx}
\usepackage{amsmath} 
\usepackage{subfigure}

\begin{document}

\newcommand{\gcm}{\text{ g cm}^{-3}}
\newcommand{\mm}{\ \mu \text{m}}
\newcommand{\cm}{\ \text{cm}}
\newcommand{\mg}{\ \mu \text{G}} 
\newcommand{\kms}{\ \text{km s}^{-1}}
\newcommand{\AU}{\ \text{AU}}
\newcommand{\V}{\ \text{Volt}}
\newcommand{\Gyr}{\ \text{Gyr}} 
\newcommand{\Myr}{\ \text{Myr}}
\newcommand{\pc}{\ \text{pc}}


\def\etal{et al.\ \rm}
\def\ba{\begin{eqnarray}}
\def\ea{\end{eqnarray}}

\title{The Dynamics of Dust Grains in the Outer Solar System}

\author{Mikhail A. Belyaev\altaffilmark{1} \& 
Roman R. Rafikov\altaffilmark{1,2}}
\altaffiltext{1}{Department of Astrophysical Sciences, 
Princeton University, Ivy Lane, Princeton, NJ 08540; 
mbelyaev@astro.princeton.edu, rrr@astro.princeton.edu}
\altaffiltext{2}{Sloan Fellow}


\begin{abstract}
We study the dynamics of large dust grains $\gtrsim 1 \mm$ with orbits
outside of the heliosphere (beyond 250 AU). 
Motion of the Solar System 
through the interstellar medium (ISM) at a velocity of $26 \kms$
subjects these particles to gas and Coulomb drag (grains are 
expected to be photoelectrically charged) as well as 
the Lorentz force and the electric force caused by the induction 
electric field. We show that to zeroth order the combined effect 
of these forces can be well described in the framework of the 
classical Stark problem: particle motion in a Keplerian potential  
subject to an additional constant force. Based on this analogy, we 
elucidate the circumstances in which the motion becomes 
unbound, and show that under local ISM conditions dust grains 
smaller than $\sim 100 \mm$ originating in the Oort Cloud 
(e.g. in collisions of comets) beyond $10^4$ AU are ejected from the
Solar System
under the action of the electric force. Orbital motion 
of larger, bound grains is described analytically using the 
orbit-averaged Hamiltonian approach and consists of orbital plane 
precession at a fixed semi-major axis, accompanied by the periodic 
variations of the inclination and eccentricity (the latter may 
approach unity in some cases). A more detailed analysis of the
combined effect of gas and Coulomb drag
shows it is possible to reduce particle semi-major axes, 
but that the degree of orbital decay is limited (a factor of several 
at best) by passages through atomic and molecular clouds, which easily
eject small particles.
\end{abstract}



\section{Introduction.}  
\label{sect:intro}


The dynamics of dust particles in the inner Solar System
has been previously addressed by many authors in different
contexts: effects of radiation
forces on circumsolar
\citep{Burns} and circumplanetary \citep{Hor2} 
grains, electromagnetic interaction of interplanetary dust 
particles with the magnetic field of the solar wind (Landgraf 
2000), grain dynamics in planetary 
magnetospheres (Horanyi 1996), and so on. 

Recent discoveries of debris disks around young stars 
have brought to light additional physical effects related
to dust dynamics such as collisions between the dust particles 
(Stark \& Kuchner 2009) and their resonant interaction with 
planetary bodies leading to asymmetries and gaps in debris 
disks (Wyatt \etal 2008). 

At the same time the dynamics of dust particles in the outer
Solar System (OSS) has received much less attention. Here the outer 
Solar System means the region of space outside of the
bow shock at 250 AU (Richardson \& Stone 2009) where unperturbed inflowing material from the interstellar medium (ISM; see Table \ref{ISMphases} for a
summary of ISM properties) undergoes a shock transition, and extending all
the way to the outer edge of the Oort Cloud of comets, roughly 
at $5\times 10^4$ AU (Fernandez 1999). The solar wind does not penetrate 
into this part of the Solar System, but the dust grains 
there move through the flow of ISM material, which
perturbs their orbits. This makes the problem of determining the
orbital evolution of grains different in the Oort Cloud as
compared to the Kuiper Belt, since the charged component of the ISM flow does not penetrate to
the Kuiper Belt; instead, the effects of the solar wind and planetary
perturbations are important there \citep{MoroMartin}. Effects of the
neutral component of the ISM flow on the Kuiper Belt dust particles 
have been considered by Scherer (2000) and P\'astor \etal (2010).

Although in this work we are concerned with the dynamics of grains in
the OSS and not their origin, we speculate that the abundance of
grains should be correlated with the spatial distribution of 
larger bodies such as
comets. This is because collisions between larger bodies create a
fragmentation cascade down to smaller sizes. At the moment our knowledge 
of spatial distribution of comets 
heavily relies on numerical simulations of Oort Cloud
formation and evolution. Such calculations typically produce 
a Cloud having both
an inner and an outer edge \citep{Kaib2,Dones}.
The location of the
outer edge in simulations is between $5\times 10^4 - 1\times
10^5$ AU, which is only a factor of a few smaller than the typical
dimension of the last
closed Hill surface \citep{Antonov}, beyond which objects are
unbound from the Solar System by the galactic tide. Another result
  from simulations \citep{Kaib2} is that the cometary density rises
  towards the inner edge, and it is thus likely that dust production is
  highest there. However, the location of
  the inner edge depends on the environment in which the Solar System
  formed. \citet{Kaib2} and \citet{Brasser} have shown that if the
  Solar System formed in a cluster, the location of the
  inner edge can vary from roughly $100 \AU$ to $3000 \AU$ depending on
  the stellar density in the cluster. Higher stellar densities help stabilize
  a planetesimal kicked out by the giant planets at a smaller value of
  the semimajor axis. For this paper, we take the inner edge
  to be at $3000 \AU$, and results by \citet{Kaib} suggest that
  most of the long period comets entering the Solar System have
  initial semimajor axes at this distance. This implies that the
  inner edge should be no further than this distance, although it
  could be closer in.

The goal of this work is to explore the dynamics of dust grains in 
the OSS by carefully analyzing different processes 
affecting their motion. Possible
observational manifestations of such grains
may provide us with information about the Oort Cloud and the
collisional processes in it. Another reason for this study is 
that dust produced in the OSS may help to understand the flow of big
interstellar dust grains
recently detected by the {\it Galileo}, {\it Ulysses}, and
{\it Cassini} satellites (Grun \etal 1994; Landgraf \etal 2000; 
Altobelli \etal 2003) and may contribute to the flux of
micro-meteoroids observed at Earth \citep{Murray,Weryk}. While 
our work was being refereed, we became aware of
the paper by P\'astor \etal (2010) which discusses similar 
processes in the Kuiper Belt. Although their work has some similarities
with our study, the methods they employed and some of their results 
are different.

This paper is organized as follows. In \S\ref{perturbations}
we analyze the importance of different forces for the dynamics
of dust grains. In \S\ref{stark} we explore the secular 
effects of these forces on grain motion in the framework
of the Stark problem. In \S\ref{sect:orbdecay} we turn to 
the decay of dust particle orbits caused by the total drag
  (combined effect of
gas and Coulomb drag), and in \S\ref{dgdapplication} we discuss
applications of our results for dust evolution in the OSS. Finally, in
\S\ref{satellites} we briefly
discuss the possibility that the large interstellar grains observed by
the satellites originate in the Oort Cloud.


\section{Forces determining grain dynamics}
\label{perturbations}


If the ISM were absent in the OSS, to a zeroth order approximation,
the dust particles
residing there would move around the barycenter of the Solar
  System (BSS) on Keplerian
orbits modified by radiation pressure. These orbits would slowly evolve 
under the action of the Poynting-Robertson (PR) drag, the galactic tide, 
and close stellar passages (Heisler \& Tremaine 1986). The Keplerian
orbital period of a body with semi-major axis $a$ is 
(neglecting radiation pressure)
\ba
T_K=2\pi\left(\frac{a^3}{GM_\odot}\right)^{1/2}=10^6\,\mbox{yr}\,
a_4^{3/2},
\label{eq:T_K}
\ea
where $a_4\equiv a/(10^4\,\text{AU})$.

The presence of the ISM flow changes this simple picture. First, 
it gives rise to gas drag on
the grains simply due to collisions of neutrals with the grain
surface. Second, grains are expected to be photoelectrically charged
to a potential
of several Volts (note that Scherer (2000) and P\'astor \etal (2010)
have considered the case of neutral grains only). If the ISM has some ionized component, this gives 
rise to the Coulomb drag, which is the electric analog of
dynamical friction \citep{galacticdynamics} and is caused by the
deflection of ions around a charged grain. Third, the magnetic 
field of the ISM also interacts with charged grains.
For small enough particles (or dense enough ISM) these forces 
can be stronger than the gravitational attraction to the BSS
making dust grains unbound (see \S\ref{particleejection}). 
In \S\S \ref{subsect:EM}-\ref{subsect:rad} we analyze the relative 
effect of these and other forces on the grain dynamics in a 
variety of circumstances. 

At present, the Solar System is moving at a velocity of
$v_w\approx 26$ km s$^{-1}$ through the warm phase of the ISM 
which is characterized by a gas number density $n({\rm H}^0)=0.2$ 
cm$^{-3}$, temperature $T_g=6300$ K, and ionization fraction 
$\chi \approx 0.25$ (Frisch \etal 2009). The strength and orientation 
of the magnetic field carried with the wind are rather uncertain, 
but a typical estimate is $B \sim 3-5\,\mu$G 
(Opher \etal 2009). As the Solar System moves through different phases of the 
ISM, the properties of the ISM wind may change quite dramatically 
compared to these numbers. Thus, we need to separately study the 
effects of the ISM flow on grains in different ISM phases. 
We assume for simplicity 
that the relative Solar System-ISM velocity stays equal to 
$v_w\approx 26$ km s$^{-1}$ at all times. Dust grains are assumed 
to be spherical and to have a density of $\rho_g=1$ g cm$^{-3}$. The grain
radius $r_g$ is a variable parameter, but in this study we will 
focus on the dynamics of rather large (by ISM standards) 
particles with $r_g\gtrsim 1\,\mu$m, because such particles have been
detected in satellite observations (Kruger \& Grun 2009).  

\begin{table}[h]
\begin{center}

\begin{tabular}{|l|c|c|c|c|c|c|c|}
\hline
ISM phase & $n_H (\text{cm}^{-3})$ & $T (K)$ & $f$ & $\chi $ & ${\cal F}$ & $s_\text{e}$ & $s_\text{i}$ \\
\hline
Coronal & $ 0.003$  & $5 \times 10^5$ & $0.5$ & $1$ & $0.5$ & $7 \times
10^{-3}$ & $0.3$ \\
Warm & $1$ & $5000$ & $0.5$ & $0.5$ & $20$ & $0.07$ & $3$  \\
Atomic cloud & $30$ & $100$ & $0.01$ & $0.02$ & $2000$ & $0.5$ & $20$ \\
Molecular cloud & $10^4$ & $25$ & $10^{-4}$ & $10^{-7}$ & $3000$ & $1$ & $40$ \\
\hline
\end{tabular}
\caption{Hydrogen number density  $n_H$, temperature $T$, filling factor 
$f$, ionization fraction $\chi$ (Draine 2010), and derived
  parameters -- drag factor ${\cal F}$ 
(equation [\ref{qeq}]), and quantities $s_e$, and $s_i$ (equation [\ref{seqn}]), which are
  proportional to the ion and
  electron Mach numbers of the grain -- for various ISM phases.} 
\label{ISMphases}
\end{center}
\end{table}


\subsection{Electromagnetic Forces}
\label{subsect:EM}

Solar motion relative to the ISM induces an electric field ${\bf
E}={\bf v}_w \times {\bf B}/c$ in the Solar reference frame, while
the magnetic field strength stays essentially the same as in the 
ISM frame. Letting $U$ be the grain potential, the electric 
${\bf F_{\text{E}}}$ and magnetic ${\bf F_{\text{B}}}$ forces on a 
grain are 
\ba
{\bf F_{\text{E}}} &=& \frac{U r_g}{c} {\bf v}_w \times {\bf B},~~~
{\bf F}_{\text{B}} = \frac{U r_g}{c} {\bf v}_g \times {\bf B},
\label{eforce}
\ea
where ${\bf v}_w$ and ${\bf v}_g$ are the wind and grain velocities in the solar frame. The
grain charge $q=Ur_g$ is taken to be constant, although we 
relax this assumption in \S\ref{starkvariations}.

Grains which are large enough to be only weakly affected by 
the ISM wind move on (perturbed) Keplerian orbits at speeds 
which are small compared to ${\bf v}_w$:
\begin{equation}
\label{relvel}
\frac{v_K}{v_w} \sim 0.01 a_4^{-1/2},
\end{equation} 
where ${\bf v}_K$ is the velocity of a 
grain moving on a circular orbit. For these grains
$|{\bf v}_g|=|{\bf v}_K|\ll v_w$ and $|{\bf F}_{B}|\ll
|{\bf F_{\text{E}}}|$ allowing us to neglect the magnetic 
force throughout this study when considering the dynamics of 
grains {\it decoupled} from the ISM gas flow. This assumption 
remains valid even for 
very small grains which get entrained in the wind
provided that these particles have just been produced in 
collisions of bigger bodies (collisional debris should move with
velocities $\lesssim v_K$ in the Solar frame) and
have not had time to get accelerated by the ISM flow to speeds 
comparable to $v_w$. Thus, in the following we will 
focus only on the electric component of the
electromagnetic force acting on grains and will mention 
the magnetic force only when 
discussing particle ejection in \S\ref{particleejection}.

Defining $\theta$ to be the angle between the magnetic field and the
wind velocity, the strength of the induction electric force $F_{\text{E}} = Ur_gv_w B
\sin\theta/c$ relative to the gravitational force $F_g=GM_\odot m_g/a^2$ is
\ba
\frac{F_{\text{E}}}{F_g} = 0.6 \, a_4^2\,
U_1\, r_{g,2}^{-2}\, B_5\, v_{w,26}\,\rho_1^{-1} \sin\theta,
\label{eq:Eforce_rel}
\ea
where $U_1\equiv U/(1\,\text{V})$, $r_{g,2}\equiv r_g/(100\,\mu$m), 
$B_5\equiv B/(5\,\mu$G), $\rho_1\equiv \rho/(1\,\text{g cm}^{-3})$, 
$v_{w,26}\equiv v_w/(26$ km s$^{-1}$), and
$m_g=(4\pi/3)\rho_g r_g^3$ is the grain mass. Because $F_{\text{E}}\propto r_g$ while $F_g\propto r_g^3$, 
the electric force becomes larger than the gravitational attraction to
the BSS for particles smaller than some
critical radius
$r_{g,\text{min}}$. Grains with $r_g<r_{g,\text{min}}$ will be 
swept up in the ISM flow and ejected from the Solar System. 
We shall find in \S\ref{particleejection} that an estimate of 
$r_{g,\text{min}}$ good to within about a factor of two can be obtained by setting $F_E/F_g = 0.25$, in
which case we find (neglecting the $\theta$-dependence)
\begin{equation}
\label{rgminE}
r_{g,\text{min}} = 150 \mm \, a_4\, U_1^{1/2}\, B_5^{1/2}\,
v_{w,26}^{1/2}\, \rho_1^{-1/2}.
\end{equation}

For particles smaller than $r_{g,\text{min}}$, we can neglect $F_g$ 
compared to electromagnetic forces.
Then in the frame of the wind, a newly created (e.g. in collisions of 
bigger grains) particle moves with speed $-{\bf v}_w$ since $v_K\ll v_w$.
This causes gyration of the particles in the frame of the wind, while
in the Solar System frame the particle will additionally experience 
an ${\bf E}\times {\bf B}$ drift with speed ${\bf v}_w$. If the
angle $\theta$ between ${v}_w$ and ${\bf B}$ is not small, one expects
small particles to get accelerated to a velocity $\sim v_w$ in the
Solar System frame on a length scale
\ba
d_{\text{coupl}} \sim R_L= \frac{m_g v_w c}{U B r_g} = 
1.3\times 10^8 \AU \, r_{g,2}^2 
\rho_1\, v_{w,26}\, U_1^{-1}\, B_5^{-1},
\label{eq:R_L}
\ea
where $R_L$ is the Larmor radius of the grain.

A notable feature of the electric force acting on grains with 
$v_g \ll v_w$ is that its magnitude
and direction are independent of either the grain's speed or its 
location, provided that the magnetic field
is homogeneous on scales comparable to the size of the Solar System. 
This significantly simplifies the analysis of the grain motion
in the electric force field as we demonstrate in \S\ref{stark}.


\subsection{Gas Drag and Coulomb Drag}
\label{gcsection}  
 
In addition to electromagnetic forces, grains in orbit around the BSS
experience gas and Coulomb drag due to the ISM. Since the
mean free path of gas molecules and ions in the ISM is much larger than
$r_g$, the total drag force on a spherical grain under the
assumption of sticking or specular reflection is given by
\citep{Baines,Drainedrag}
\begin{eqnarray}
\label{drag}
F_{\text{drag}} = \pi r_g^2 P {\cal F}, 
\end{eqnarray}
where
\begin{eqnarray}
P &\equiv & n k T 
\end{eqnarray}
is the full ISM pressure ($n$ is the particle number density, $T$ is
the temperature) and 
\begin{eqnarray}
\label{qeq}
{\cal F} &\equiv &  \frac{2}{n} \sum_i n_i\left\{G_0(s_i) + 
\frac{1}{2}z_i^2 \phi^2
  \ln\left[1 + \left(\frac{\Lambda}{z_i}\right)^2\right] G_2(s_i)\right\}, \\
\label{eq:G0}
G_0(s) &\equiv & \left(s^2 + 1 - \frac{1}{4s^2}\right) \text{erf}(s) +
  \frac{1}{\sqrt{\pi}} \left(s + \frac{1}{2s} \right) e^{-s^2}, \\
G_2(s) &\equiv & \frac{\text{erf}(s)}{s^2} - \frac{2}{s \sqrt{\pi}}
  e^{-s^2}, \\
\label{seqn}
s_i &\equiv& \left( \frac{m_i v^2}{2kT} \right)^{1/2}, \\
\phi &\equiv& \frac{eU}{kT}, \\
\Lambda &\equiv& \frac{3}{2r_ge|\phi|}\left( \frac{kT}{\pi n_e}
  \right)^{1/2}. 
\label{eq:Lambda}
\end{eqnarray}
The sum in equation (\ref{qeq}) runs over all particle species $i$, and
the first and second terms are due to the gas drag and Coulomb
drag respectively. The term $\frac{1}{2}\ln\left[1+(\Lambda/z_i)^2\right]$ 
in equation (\ref{qeq}) is a generalization of the usual expression for
the Coulomb logarithm $\ln(\Lambda/z_i)$ \citep{galacticdynamics}, and
$z_i$ is the charge of species $i$. This
is necessary because for the grain sizes we consider, we sometimes 
find $\Lambda \lesssim 1$. Table \ref{ISMphases} shows the
values of the dimensionless factors $s$ and ${\cal F}$ for ions and 
electrons in various phases of the ISM. Note that 
according to equation (\ref{eq:G0}) even for neutral grains ($\phi=0$), the
drag force does not scale quadratically with grain velocity as was
assumed in Scherer (2000) and P\'astor \etal (2010); such a scaling 
is only valid for $s\gg 1$, which is not always true, see Table
\ref{ISMphases}.

We estimate the total drag force on the particles by 
setting\footnote{In \S\ref{dragseparation} we abandon this 
simplifying assumption. } $v= v_w$,
which is justified as long as $v_g \ll v_w$ (certainly true for large 
grains, see \S\ref{subsect:EM}):
\begin{equation}
\frac{F_{\text{drag}}}{F_g} = 0.4 \, a_4^2 \, \rho_1^{-1} \,
r_{g,2}^{-1} \, {\cal F}_{20} \, P_1,
\label{eq:Fdrag_rel}
\end{equation}
where ${\cal F}_{20} = {\cal F}/20$ and $P_1 = nkT/(\text{eV} \cm^{-3})$.
Unlike the electric force which scales as $\propto r_g$, the total drag force
scales as $\propto r_g^2$, meaning that there is a critical particle
radius
\begin{equation}
r_{g,\text{crit}} = 140 \mm \, U_1\, v_{w,26}\, B_5\, {\cal F}_{20}^{-1} P_1^{-1}
\end{equation}
for which the electric force equals the total drag force. 

Additionally, if the total drag force dominates the electric force, then the
minimum size of a bound particle obtained as before by setting 
$F_{\text{drag}}/F_g=0.25$ is
\begin{eqnarray}
r_{g,\text{min}} &=& 160 \mm \, a_4^2 \, \rho_1^{-1} \,  {\cal F}_{20} P_1.
\label{eq:bound_drag}
\end{eqnarray}
The coupling distance to the ISM flow $d_{\text{coupl}}=m_g v_w^2/
(2F_{\text{drag}})$, defined as the length scale over which the work 
done by the total drag force (evaluated at $v=v_w$) is equal to the 
kinetic energy of the grain moving at speed $v_w$, is given by
\begin{eqnarray}
d_{\text{coupl}} &=& 9 \times 10^7 \AU \, r_{g,2} \, \rho_1 \,
v_{w,26}^2 \, {\cal F}_{20}^{-1} P_1^{-1}.
\label{eq:d_coupl}
\end{eqnarray}

Since the different phases of the ISM are in rough pressure 
equilibrium one might expect that $F_{\text{drag}}$ should be
comparable in all phases (one notable exception being the molecular
clouds in which $P$ is higher than in the average ISM owing to their
self-gravity) as $F_{\text{drag}}$ is directly
related to gas pressure $P$ (equation [\ref{drag}]). In reality we find that $F_{\text{drag}}$
varies greatly depending on the phase of the ISM, because the dimensionless
factor ${\cal F}$ varies by orders of magnitude reflecting different 
ionization levels and Mach numbers in different phases, even though
the pressure is approximately constant between different phases, see Table 
\ref{ISMphases}. For this reason, the values of $r_{g,\text{crit}}$ and
$r_{g,\text{min}}$ also vary dramatically in different environments, 
see Figures \ref{radiusradius} and \ref{ISMejection}.

\subsubsection{Two dynamically distinct contributions 
to the total drag}
\label{dragseparation}

Dust grains with sizes $\gtrsim r_{g,\text{min}}$ feel gas and Coulomb
drag 
only as a perturbation on top of the dominant gravitational force. To
zeroth order, they continue moving on Keplerian orbits
(albeit with time-varying orbital elements) and the relative 
speed between these grains and the ISM flow is ${\bf v}={\bf v}_w+
{\bf v}_K$, where ${\bf v}_K$ is the Keplerian velocity of the grain. 
Assuming for simplicity that the ISM and grain properties are
constant as the grain moves on its orbit, the total drag force can 
be split as
\begin{eqnarray}
{\bf F}_{\rm drag} = - F(v) \frac{{\bf v}}{v}= 
{\bf F}_{{\rm drag},0}+\Delta{\bf F}_{\rm drag} +
  \mathcal{O}\left[\left(\frac{v_K}{v_w}\right)^2\right],
  \label{masterdrag}
\end{eqnarray}
where
\ba
{\bf F}_{{\rm drag},0} &=&-F(v_w) \frac{{\bf v}_w}{v_w},
\label{eq:F_0}\\
\Delta{\bf F}_{\rm drag} &=& - F(v_w)\frac{{\bf v}_K}{v_w} + \left[
  -\frac{d F}{d v} \Big |_{v=v_w} v_w + F(v_w) \right]
  \frac{{\bf v}_w \cdot {\bf v}_K}{v_w^2}\frac{{\bf v}_w}{v_w},
\label{eq:DeltaF}
\ea
and $F(v)$ is the amplitude of the total drag force. 
The term ${\bf F}_{{\rm drag},0}$ is the zeroth order component 
which one obtains by setting the orbital velocity to zero, and 
$\Delta {\bf F}_{\rm drag}$ is the correction which is linear 
in $v_K/v_w$. Because ${\bf F}_{{\rm drag},0}$ 
is independent of ${\bf v}_K$ and the position of the dust 
particle, its effect on grain dynamics is analogous to that of
the electric force ${\bf F}_{\text{E}}$, see \S\ref{subsect:EM}.
We study the combined effect of these conservative forces
on the grain motion in \S\ref{stark}.

The {\it differential} drag force $\Delta {\bf F}_{\rm drag}$ (due to
both Coulomb and gas drag)
depends explicitly on ${\bf v}_K$ which makes its action similar to 
that of a frictional force. We analyze the effect that this force 
has on the semi-major axis of dust grains in \S\ref{sect:orbdecay}.


\subsection{The Galactic Tide}
\label{galactictide}

The importance of the galactic tide for the dynamics of bodies in the 
OSS has been recognized for a long time. 
Using results of Heisler \& Tremaine (1986) we find the ratio of the 
tidal force $F_\text{tide} = 16 \pi r_g^3 \rho G \rho_0 z/3$ to the Solar
gravitational force to be
\begin{equation}
\frac{F_\text{tide}}{F_g} = 2.6 \times 10^{-4} \,
  z_4 \, a_4^2 \left( \frac{\rho_0}{0.185 \ M_{\sun} \, \text{pc}^{-3}}
  \right),
\end{equation} 
where $z$ is the vertical distance between the BSS and the object 
(measured in the direction perpendicular to the galactic plane), 
$z_4\equiv z/(10^4\,\text{AU})$, and
$\rho_0 = 0.185 \ M_{\sun} \, \text{pc}^{-3}$ is the average stellar density in the
galactic disk near the Sun \citep{Bahcall}.

If all other perturbations
were negligible, the galactic tide would be important beyond $a \sim
3000$ AU, since beyond this distance, the time for the orbital elements
to cycle back to their initial values is shorter than the age of the
Solar System \citep{Heisler}. If gas drag, Coulomb drag, and the electric force
are considered, though, the picture changes. The tidal force
is proportional to $z$ and scales as $\propto r_g^3$ in the particle radius,
whereas the electric and drag forces are independent of $z$ and
scale as $\propto r_g$ and $\propto r_g^2$
respectively. Thus, the tidal force dominates over the electric and
drag forces {\it only} for large enough $r_g$ and $z$, and such particles are
dynamically similar to comets.  Smaller particles, or those with
smaller values of $z$ are dynamically similar to grains, and the
evolution of their
orbital elements is governed by the electric and total drag forces.

\begin{figure}[!p] 
  \centering
  \includegraphics[width=.8 \textwidth]{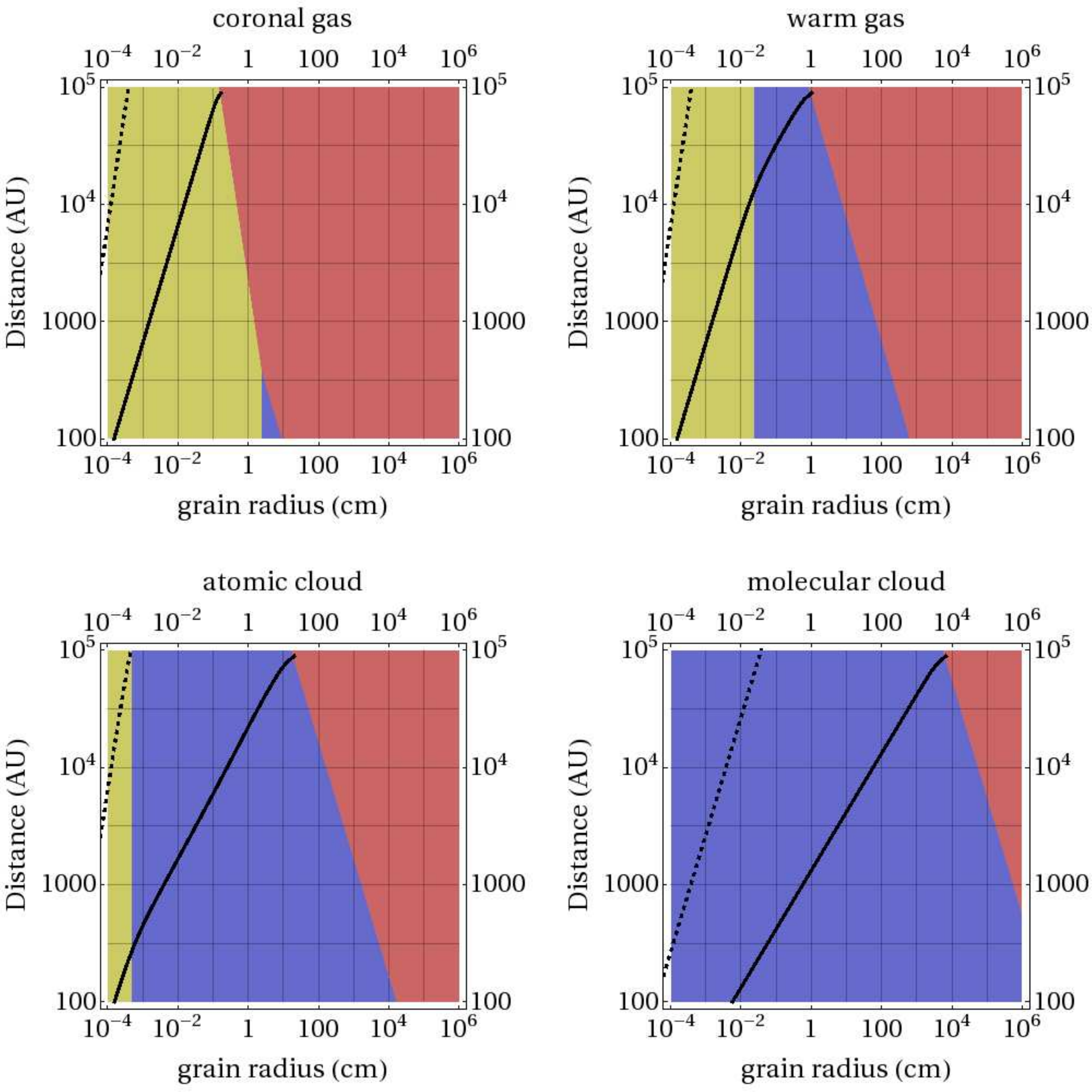}
  \caption{Map of regions in particle radius - orbital distance
    parameter space in which different perturbing forces dominate as
    shown by the different colors: {\it yellow} - electric force, 
{\it blue} - gas drag 
  and Coulomb drag combined, {\it red} - galactic tide. Different panels
    correspond to different ISM phases as labeled on the figure. The
    solid line gives the
particle size below which ejection occurs (see \S\ref{particleejection}) 
and the dashed line
  gives the condition $a=d_\text{coupl}$ (equations [\ref{eq:R_L}] and
    [\ref{eq:d_coupl}]). We assume that
  the grains have a density of $1 \gcm$ and are at a potential of $1
  \V$. The magnetic field has a strength of $5 \mg$, and the ISM
  properties are the same as those listed in Table \ref{ISMphases}.} 
  \label{radiusradius}
\end{figure}

The particle size at which $F_{\text{tide}}$ is comparable to the
electric force is
\begin{equation}
r_{g,\text{crit}} = 0.5 \cm \, z_4^{-1/2} \, \rho_1^{-1/2} \, U_1^{1/2} \, B_5^{1/2}
  \, v_{w,26}^{1/2}
  \left( \frac{\rho_0}{0.185 \ M_{\sun} \, \text{pc}^{-3}} \right)^{-1/2}, 
\label{eq:tideelectric}
\end{equation}
and the particle size at which $F_{\text{tide}}$ equals the total 
drag force is
\begin{equation}
r_{g,\text{crit}} = 8 \cm \, z_4^{-1} \, \rho_1^{-1} \, {\cal
  F}_{20} P_1 \left( \frac{\rho_0}{0.185 \ M_{\sun} \, \text{pc}^{-3}}
  \right)^{-1}.
\label{eq:tidedrag}
\end{equation}
Figure \ref{radiusradius} illustrates the cutoff between particles
which are dynamically similar to grains and those that are dynamically
similar to comets for different ISM phases. 

In this study we will not consider the effects of the galactic tide, 
since they have already been investigated by e.g. Heisler \& Tremaine 
(1986). Instead, we will limit our discussion of the non-gravitational 
forces acting on dust particles with sizes smaller than $r_{g,\text{crit}}$
given in equations (\ref{eq:tideelectric}) and (\ref{eq:tidedrag}).

\subsection{Radiation Forces}
\label{subsect:rad}

There are several ways in which radiation affects the grain motion. 
Solar radiation exerts the radiation pressure force on dust particles
\begin{equation}
F_{\text{rad}} = \frac{L_{\odot} r_g^2}{4 a^2 c}\langle Q\rangle,
\end{equation}
where $c$ is the speed of light and $\langle Q\rangle$ is a 
frequency-averaged radiation absorption 
and scattering coefficient (Burns \etal 1979),
which equals 1 in the regime of geometrical optics. 
Because $F_\text{rad}$ is directed radially and scales in the same
way as gravity, it simply modifies ${\bf F}_g$ to be ${\bf F}_g(1-\beta)$
where $\beta\equiv F_\text{rad}/F_g$. For particles larger than $1 \mm$ 
in size $\beta$ is small (Burns \etal 1979), and as we show later, 
particles smaller than $1\,\mu$m in size are ejected by the electric and 
total drag forces outside of the heliosphere (at $a>250$ AU). For that 
reason we simply disregard Solar radiation pressure in this
study (if needed the contribution of the radiation pressure
can be easily accounted for by redefining $M_\odot\to M_\odot
(1-\beta)$ in all equations).

Poynting-Robertson drag is unimportant beyond $250 \AU$ 
since the decay time for a circular orbit is 
\begin{eqnarray}
&& \tau_{PR} = \frac{16 \pi r_g \rho a^2 c^2}{3 L_{\odot} \langle Q \rangle} 
\label{eq:PR}\\
&&= 2.8 \times 10^{4} \text{ Gyr} \, r_{g,2} \rho_1 a_4^2 \langle
Q\rangle^{-1}  \nonumber
\end{eqnarray} 
Although this is shorter than the age of the Solar
System for $1 \mm$ grains inside of $10^3 \AU$, these 
grains are ejected from the Solar System on significantly 
shorter timescales by other non-gravitational forces as
we demonstrate in \S\ref{dgdapplication}. 

The anisotropy of background starlight exerts a force 
$F_{\text{anis}} =\Delta \pi r_g^2 u \langle Q \rangle$, where $u =
10^{-12}$ erg cm$^{-3}$ \citep{Drainebook} is the
energy density of the background starlight and
$\Delta= 0.1$ \citep{Weingartner} is a parameter quantifying the degree
of anisotropy. We estimate that
\begin{eqnarray}
\frac{F_\text{anis}}{F_g} =  1.3 \times 10^{-3} \, r_{g,2}^{-1} \, \rho_1^{-1} \, a_4^2
\left(\frac{\Delta}{0.1}\right) \left(\frac{u}{10^{-12}
  \ \text{erg} \ \text{cm}^{-3}}\right) \langle Q \rangle, 
\end{eqnarray}
and using equations (\ref{eq:Eforce_rel}) and 
(\ref{eq:Fdrag_rel}) we find that 
$F_{\text{anis}}$ is sub-dominant compared to either the total drag,
the electric force, or both in all ISM phases. 

Finally, there is also a non-conservative drag force that arises due 
to the redshifting and blueshifting of the background starlight as 
the dust grain orbits the BSS, similar to the differential drag 
described in \S\ref{dragseparation}. However the magnitude of this 
force is $\sim F_{\text{anis}} \Delta^{-1}\left(v_K/c \right)$
and this is always much smaller than 
$|\Delta {\bf F}_{\rm drag}|\sim \left(v_K/v_w\right)F_{\text{drag}}$.
Thus, for the problem at hand we can safely neglect any radiation 
forces on dust grains.


\section{The Stark Problem}
\label{stark}


In the approximation that the ISM properties and $\bf{v}_w$ are held
constant and $v_w \gg v_K$,
the electric force and the velocity-independent total drag force 
$F_{{\rm drag},0}$ do not depend on the orbital parameters 
and are constant over an orbit. This reduces the problem of solving 
the grain dynamics to the classical Stark problem of 
motion in a Keplerian potential subject to an extra force ${\bf S}$ 
which is constant in magnitude and direction. In our case,
\ba
{\bf S}={\bf F}_E+{\bf F}_{{\rm drag},0},
\label{eq:S}
\ea
has both a component ${\bf F}_{{\rm drag},0}$ parallel to 
${\bf v}_w$ and a component ${\bf F}_E$ orthogonal to 
${\bf v}_w$. Thus, in general
${\bf S}$ is oriented arbitrarily with respect to ${\bf v}_w$.

The classical Stark problem in its most general setting, including the case 
of $S\equiv |{\bf S}|\gtrsim F_g$, has been previously explored 
analytically. It has been shown to be separable in parabolic coordinates by \citet{LL}
and by Banks \& Leopold (1978), who studied the 
ionization of highly excited atoms by an electric field. Kirchgraber (1970) has shown that it is
possible to solve it by using the 
Kustaanheimo-Stiefel (KS) transformation (Kustaanheimo \& Stiefel 1965), 
which maps out the three-dimensional
Keplerian problem into the four-dimensional harmonic oscillator problem. 
 Unfortunately,
the results of these studies are expressed in terms of integrals of 
Jacobian elliptic functions, the special functions inverse 
to the elliptic integrals, and are thus difficult to
analyze. Nevertheless, we will use the existing analyses of
the Stark problem in \S\ref{particleejection} to obtain an accurate description of the conditions 
under which particle ejection occurs.

In most of this study, however, we are interested in the motion of
grains large enough for the Stark force, equation (\ref{eq:S}), to be considered
a perturbation on top of $F_g$. We can quantify this condition as
\ba
\alpha\equiv\frac{Sa^2}{GM_\odot m_g} \ll 1,
\label{eq:alph}
\ea
where $a$ is the instantaneous value of the semi-major axis. 
In this limit, we can use standard methods of perturbation theory 
to investigate the dynamics of dust particles. Previously, \citet{Mignard} have used
osculating elements \citep{Burns,MD} to solve the problem of a constant force perturbing a test particle orbiting a central mass which is stationary in a rotating reference frame. This problem is relevant to the study of a particle orbiting a planet subjected to radiation pressure from the Sun, and it is identical to the Stark problem if the rotation rate of the reference frame is set to zero.

We will approach the perturbative Stark problem ($\alpha\ll 1$)
in two steps. First, in \S\ref{subsect:Stark_planar} we 
consider the simplified case of the planar 
Stark problem in which ${\bf S}$ lies in the plane of the particle 
orbit using osculating 
elements. This provides us with a simple qualitative
picture of the secular evolution under the action of Stark force.
We then adopt the orbit-averaged Hamiltonian formalism to study the more 
general Stark problem in which ${\bf S}$ can have any orientation 
with respect to the orbital plane. We also compare our results 
to numerical orbit integrations using an integrator described in 
Appendix \ref{WHintegrator}.


\subsection{The Planar Stark Problem: perturbative approach.}
\label{subsect:Stark_planar}

In the case of the Stark force lying in the plane of
the orbit, the motion is restricted to this plane, and the
inclination of the orbit $i=90^\circ$ does not change. We choose 
a coordinate frame in this plane so that the $x$-axis is aligned 
with ${\bf S}$ and count the longitude of pericenter $\omega$ 
from this direction. The orbit averaged equations for the evolution
of the osculating elements \citep{Burns,MD} then become
\begin{eqnarray}
  \label{dadt}
  \left \langle \frac{da}{dt} \right \rangle &=& 0, \\
  \label{dedt}
  \left \langle \frac{de}{dt}  \right \rangle &=& -\frac{3}{2}
  \frac{JS}{GM_\odot m_g} \sin(\omega), \\
  \label{dwdt}
  \left \langle \frac{d\omega}{dt} \right \rangle &=& -\frac{3}{2}
  \frac{JS}{GM_\odot m_g e} \cos(\omega), 
\end{eqnarray}
where $J$ is the specific angular momentum and $e$ is the orbital eccentricity. 
Both $J$, $e$, and $\omega$ vary in time under the action of Stark 
force, while $a$ does not vary on average. 

\begin{figure} [h] 
  \centering
  \subfigure{\includegraphics[width=.45\textwidth]{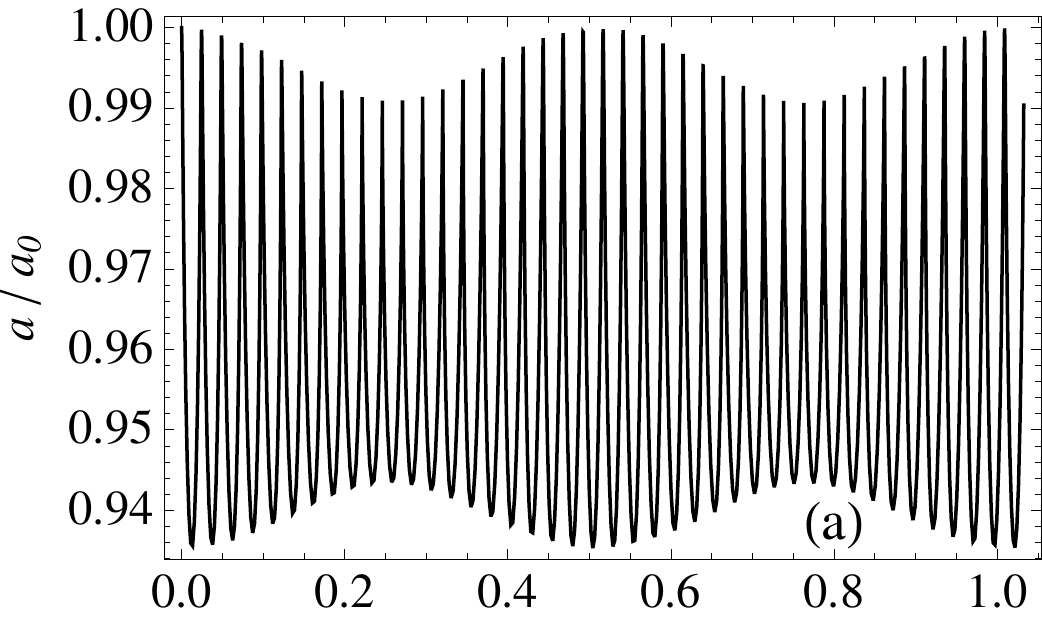}}
  \subfigure{\includegraphics[width=.45\textwidth]{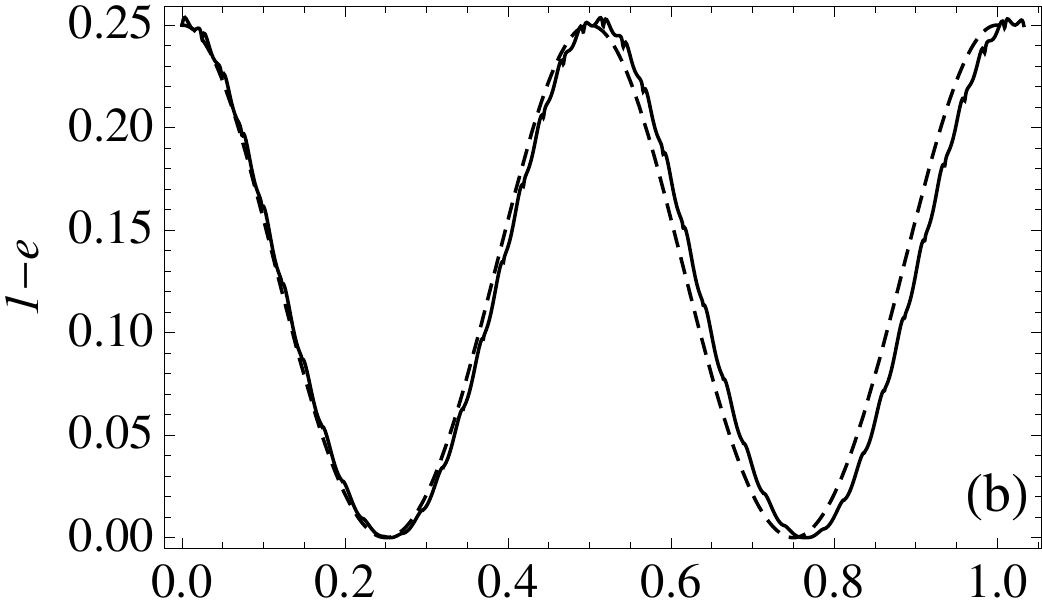}} 
  \subfigure{\includegraphics[width=.45\textwidth]{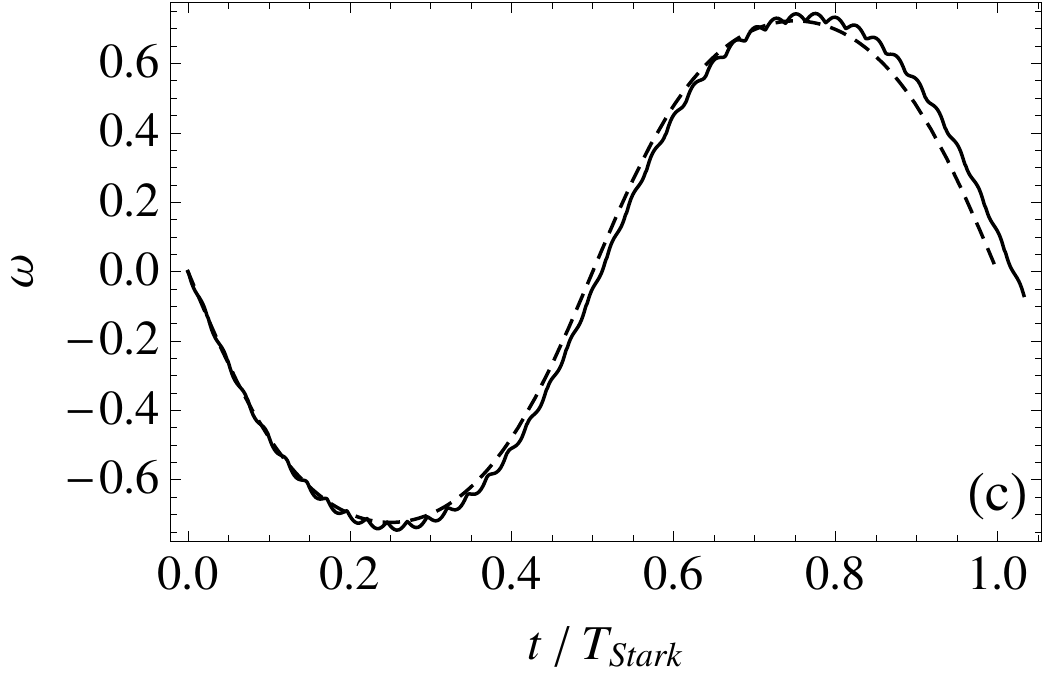}}
  \subfigure{\includegraphics[width=.45\textwidth]{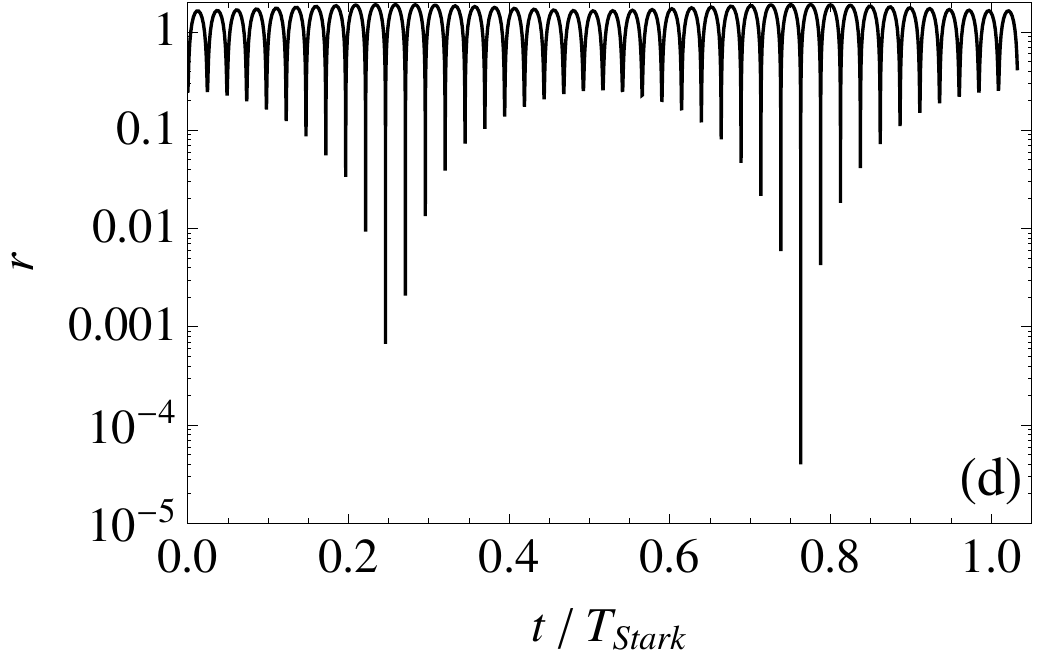}}
  \caption{Evolution of (a) semi-major axis,
  (b) eccentricity, (c) longitude of pericenter, and (d) radius over a Stark
  period for an orbit with $T_\text{stark}/T_K =
  37$ ($\alpha$ = 0.018), $\Lambda = e(0) =0.75$ and $t_0 = \omega(0) = 0$. Solid lines
  correspond to solutions obtained from numerical integration, and
  dashed lines are the analytic solutions from equations (\ref{ecosw}) and
  (\ref{esinw}). The eccentricity tends to unity when the angular
  momentum vector changes sign, and when this happens the particle can
  pass very close to the origin.}
  \label{starkelems}
\end{figure}

We can integrate equations (\ref{dedt}) and (\ref{dwdt}) directly to obtain
\begin{eqnarray}
\label{ecosw}
  e \cos\omega &=& \Lambda, \\
\label{esinw}
  e \sin \omega &=& -\sqrt{1-\Lambda^2}\sin \left( 2\pi \frac{t-t_0}{T_\text{stark}}
\right),
\end{eqnarray}
where $\Lambda$ is the value of $ e \cos\omega$ at time $t=0$,
$t_0$ is the other integration constant, and
\ba
T_\text{stark} \equiv \frac{4 \pi m_g}{3 S}
\sqrt{\frac{GM_\odot}{a}}= 0.67 \Myr \, a_4^{3/2} \, \alpha^{-1} 
\label{starkperiod}
\ea
is the timescale for the orbital elements to return to their original
values (a Stark cycle), with $\alpha$ defined in equation
(\ref{eq:alph}). The ratio of the Stark period to an orbital period
is $T_\text{stark}/T_K = (2/3)\alpha^{-1}$. Equation (\ref{ecosw}) 
agrees with the results of P\'astor \etal (2010), and the Stark period was 
previously obtained by \citet{Mignard}.

In Figure \ref{starkelems} we show
the evolution of the orbital elements over a timescale
$T_\text{stark}/4$, which corresponds to $\alpha = 0.018$. The 
agreement between the analytical theory (equations 
[\ref{ecosw}] and [\ref{esinw}]) and numerical results is good. Note
that in accordance with equation (\ref{dadt}),
the semi-major axis of the orbit stays constant because the constant 
perturbing force does no work over a closed orbit. However, 
on a time scale $T_K$, $a$ still experiences small oscillations with amplitude
\begin{equation}
\frac{\Delta a}{a} \sim 4 \alpha,
\label{aorbvar}
\end{equation}
which are due to the work done by the Stark force in the course of orbital
motion.

The eccentricity of the orbit varies through the 
Stark cycle from $\Lambda$ to $1$, thus allowing very close 
approaches of particles to the Sun. As we will see in the 
next section, such high values of $e$ are a peculiarity of the planar Stark
problem, but the periodic variations of $e$ and $\omega$ are 
generic.


\subsection{The General Stark Problem: perturbative approach}
\label{subsect:Stark_3D}

Following the procedure outlined in \citet{Heisler}, we treat 
the general Stark problem of an arbitrarily oriented Stark force ${\bf S}$ 
by orbit-averaging the particle Hamiltonian and determining the 
integrals of motion. This procedure is different from the
approach of P\'astor \etal (2010) and \citet{Mignard} who used orbit averaged 
equations for the evolution of the osculating elements 
\citep{Burns,MD} to treat the general Stark problem.

If we align the $z$-axis with ${\bf S}$, the 
Hamiltonian per unit mass becomes
\begin{equation}
H = \frac{v^2}{2} - \frac{GM_\odot}{r} - \frac{S}{m_g}z.
\end{equation}
Using the relations $z= r \sin i \sin(f+\omega)$ and 
$r=a(1-e^2)/(1+e\cos f)$, where $f$ is the true anomaly (varying on
an orbital timescale), we can rewrite this expression as 
\begin{equation}
H = -\frac{GM_\odot}{2a} + \frac{Sa(1-e^2) \sin(f+\omega)\sin i}{m_g(1 +
  e\cos f)}.
\end{equation}
When $\alpha \ll 1$ the orbital elements vary on a timescale $T_\text{stark}$, 
which is much longer than $T_K$. Averaging the Hamiltonian over a
closed Keplerian orbit we obtain
\begin{equation}
\label{Hav}
H_{\rm av} = -\frac{GM_\odot}{2a} - 
\frac{3S}{2m_g}a\, e\,\sin i \,\sin \omega .
\end{equation}
By introducing the Delaunay angle-action variables 
(Binney \& Tremaine 2008)
\begin{eqnarray}
l &,& L = \sqrt{GM_\odot a} \\
\omega &,& J = \sqrt{GM_\odot a(1-e^2)} \\
\Omega &,& J_z = J\cos i
\end{eqnarray}
($l$ is the mean anomaly, $\Omega$ is the longitude of ascending node) 
we rewrite the time-averaged Hamiltonian as
\begin{equation}
\label{Havdel}
H_{\text{av}} = -\frac{(GM_\odot)^2}{2 L^2} -
\frac{3}{2}\frac{S L^2}{GM_\odot m_g}
\sqrt{1-\frac{J^2}{L^2}}\sqrt{1-\frac{J_z^2}{J^2}}
\sin \omega.
\end{equation}
Because the Delaunay variables are canonically conjugate and the
Hamiltonian is independent of $l$, $\Omega$, and $t$, we immediately see that
$L$, $J_z$, and $H_\text{av}$ must be conserved, implying that the
semi-major axis, the $z$-component of the angular momentum, and total 
energy do not change on average. The total angular momentum $J$ is not
conserved, so that $e$, $i$, $\omega$ and $\Omega$ must vary under
the action of Stark force.

Introducing dimensionless variables 
\ba
K = \frac{J}{L},~~~~K_z = \frac{|J_z|}{L},
\label{eq:dimlessmomenta}
\ea
we find from equation (\ref{Havdel}) that
\begin{equation}
(\Delta \tilde H)^2 = (1-K^2) \left( 1-\frac{K_z^2}{K^2} \right) \sin^2 \omega,
\label{eq:C^2}
\end{equation}
is an integral of motion which we will use from now on instead of
$H_{\rm av}$. It is easy to see that $0\leq \Delta\tilde H\leq
1-K_z$. 

\begin{figure}[h] 
  \centering
  \subfigure{\label{contours}
  \includegraphics[width=.45\textwidth]{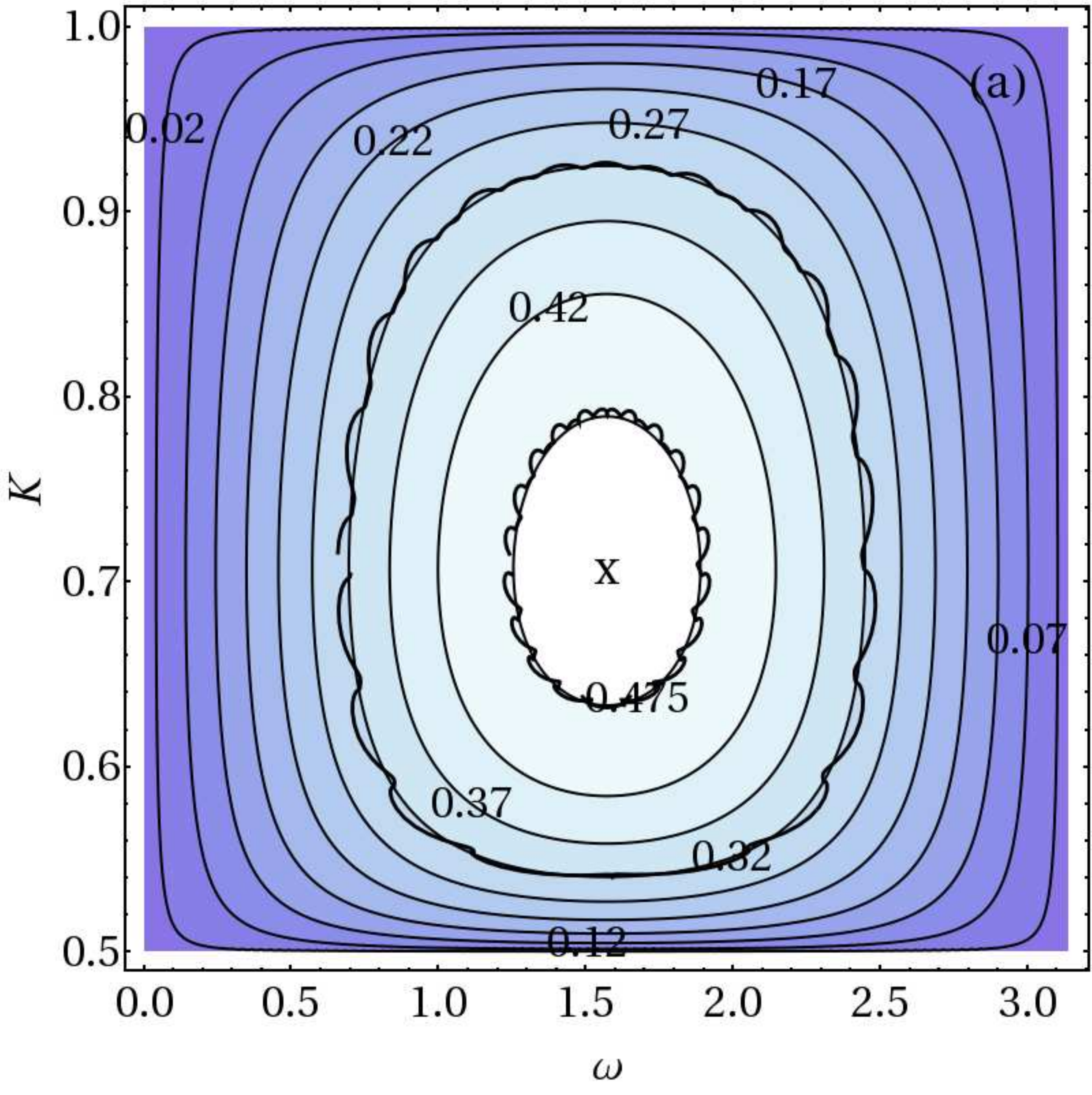}}
  \subfigure{\label{stationaryorbit}
  \includegraphics[width=.45\textwidth]{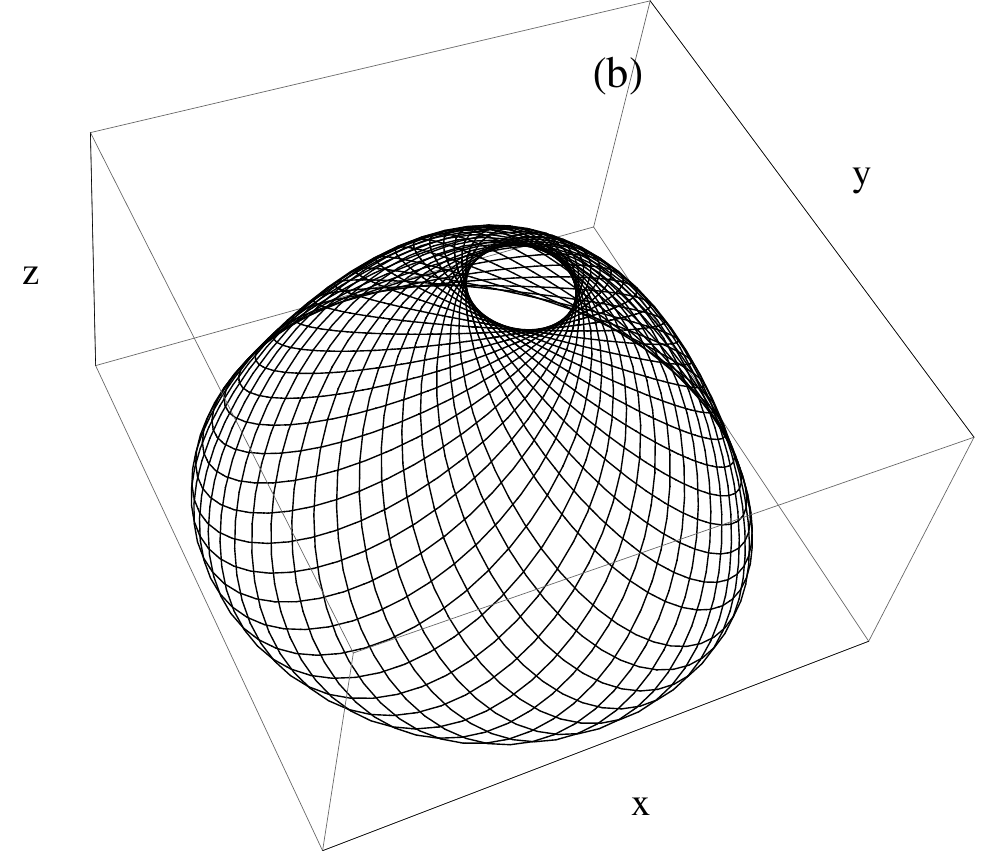}}
  \caption{(a) Contours of constant $\Delta \tilde H$ (labeled by the
  value of $\Delta \tilde H$) for $K_z = 0.5$
  with numerically integrated orbits over-plotted (both orbits have
  $\alpha = 0.02)$. The orbits do not exactly
  follow a contour due to our orbit-averaging of the Hamiltonian. (b)
  Trajectory of the stationary orbit ($\alpha = 0.02$, $K_z = 0.5$) for
  which $a$, $e$, $i$, and $\omega$ all remain constant, while
  $\Omega$ increases at a constant rate. The stationary orbit is
  marked by a cross in panel (a).}
\end{figure}

Conservation of $\Delta \tilde H$ and $K_z$ allows us to directly relate $K$ 
to $\omega$. For a given value of $K_z$ in $K-\omega$ space, the particle is confined to move on
contours of constant $\Delta \tilde H$. We have plotted the
contours for $K_z = 0.5$ in Figure \ref{contours} along with the 
numerically integrated particle trajectory in these coordinates. 
This figure clearly shows the existence of a stationary solution when
$\Delta \tilde H$ takes its
maximum value of $1-K_z$ with both $K= \sqrt{K_z}$ 
and $\omega=\pi/2$ constant in time. The three-dimensional trajectory
of this orbit is shown in 
Figure \ref{stationaryorbit}, and it is obvious that $\Omega$ is the
only orbital element of the stationary orbit that varies with time
(equation [\ref{eq:OofK}]).

We now determine the explicit time dependence of $e$, $i$, $\omega$ 
and $\Omega$ in the general case for arbitrary $J_z,L, \text{ and }
\Delta\tilde H$.
The evolution of $\omega$ is governed by the equation
\begin{eqnarray}
\frac{d\omega}{dt}=\frac{\partial H_\text{av}}{\partial J}=-\frac{3}{2}
\frac{SL}{GM_\odot m_g}\frac{K_z^2-K^4}{K^2\sqrt{(1-K^2)(K^2-K_z^2)}}
\sin\omega,
\label{eq:omegadot}
\end{eqnarray}
where $H_\text{av}$ is the Hamiltonian in equation (\ref{Havdel})).
Using equation (\ref{eq:C^2}) to express $\sin\omega$ and $d\omega/dK$
as functions of $K$ we derive the following equation for the evolution of $K$:
\begin{eqnarray}
\frac{dK}{dt}=\frac{3}{2}\frac{SL}{GM_\odot m_g}\frac{\sqrt{(K_\text{max}^2-K^2)
(K^2-K_{\text{min}}^2)}}{K},
\label{eq:Kdot}
\end{eqnarray}
where 
\begin{equation}
K^2_{\{^\text{max}_{\text{min}}\}}=\frac{1+K_z^2-(\Delta\tilde H)^2}{2}\pm 
\left[\left(\frac{1+K_z^2-(\Delta\tilde H)^2}{2}\right)^2-K_z^2\right]^{1/2}
\label{eq:Kmaxmin}
\end{equation}
are the maximum and minimum possible values of $K^2$ for given 
$K_z$ and $\Delta\tilde H$. Integrating equation (\ref{eq:Kdot}) one finds
\begin{eqnarray}
K^2(t)=\frac{K_{\text{min}}^2+K_\text{max}^2}{2}+\frac{K_\text{max}^2-K_{\text{min}}^2}{2}
\sin\left(4\pi\frac{t-t_0}{T_\text{stark}}\right),
\label{eq:Koft}
\end{eqnarray}
where $t_0$ is the integration constant. It immediately follows from 
this solution that the timescale for the orbit to go once around a 
contour in $K-\omega$ space (Figure \ref{contours}) is $T_\text{stark}/2$
independent of $K_z$ and $\Delta\tilde H$.

Since $e^2(t)=1-K^2(t)$ and 
$\cos i=K_z/K(t)$ (equation [\ref{eq:dimlessmomenta}]), the 
solution, equation (\ref{eq:Koft}), immediately gives us the time dependence of 
$e$ and $i$. Also, equation (\ref{eq:C^2}) connecting $\omega$ and $K$ 
enables us to determine explicitly the evolution of $\omega(t)$. 

Finally, to determine the evolution of $\Omega$, we use the equation
\begin{eqnarray}
\frac{d\Omega}{dt}=\frac{\partial H_{\rm av}}{\partial J_z}=\frac{3}{2}
\frac{SL}{GM_\odot m_g}\frac{\sqrt{1-K^2}}{\sqrt{K^2-K_z^2}}
\frac{K_z}{K}\sin\omega.
\label{eq:Omega_dot}
\end{eqnarray}
Equation (\ref{eq:Kdot}) allows us to convert this expression into 
an equation for $\Omega(K)$ with the solution 
\begin{equation}
\Omega(K) = \arctan \left( \sqrt{
\frac{(K^2-K_{\text{min}}^2)(K_\text{max}^2-K_z^2)}
{(K_\text{max}^2-K^2)(K_{\text{min}}^2-K_z^2)}} \right).
\label{eq:OofK}
\end{equation}
Since the argument of arctangent in the numerator of equation
(\ref{eq:OofK}) vanishes at $K =
K_{\text{min}}$ and the denominator equals zero for $K = K_\text{max}$, we have
$\Omega(K_\text{max}) - \Omega(K_{\text{min}}) = \pi/2$. Because $K$ goes from
$K_{\text{min}}$ to $K_\text{max}$ on a timescale of $T_{\text{stark}}/4$,
$\Omega$ goes from 0 to $2 \pi$ on a timescale of
$T_{\text{stark}}$. Thus, all of the orbital elements return to their
original values after a time $T_\text{stark}$ in the general
case just as in the planar case. The stationary solution has $\Omega$
linearly growing in time, which is most easily seen directly from 
equation (\ref{eq:Omega_dot}) but can also be derived by taking the
limit $\Delta \tilde H\to 1-K_z$, $K\to \sqrt{K_z}$ in equation
(\ref{eq:OofK}).

These results allow us to put constraints on the possible values of 
eccentricity and inclination that an orbit attains during its Stark cycle.
The minimum and maximum values of $K$ are $K_{\text{min}}$ and
$K_\text{max}$, and $\cos i$ oscillates 
between $K_z/K_\text{max}$ and $K_z/K_{\text{min}}$, while the
eccentricity $e$ varies between $\left(1-K_\text{max}^2\right)^{1/2}$
and $\left(1-K_{\text{min}}^2\right)^{1/2}$. Since $K_z \le K_\text{min}$,
the eccentricity only evolves to unity if the Stark vector is in the
plane of the orbit ($K_z = 0$).

\citet{Mignard} have previously found implicit formulas for the time dependence of the orbital elements. Their results also apply to a rotating reference frame,
but our formulae are more straightforward to use in the case of the Stark problem,
since they give a simple, explicit prescription for the evolution of the orbital elements.

\begin{figure}[h] 
  \centering
  \includegraphics[width=.9 \textwidth]{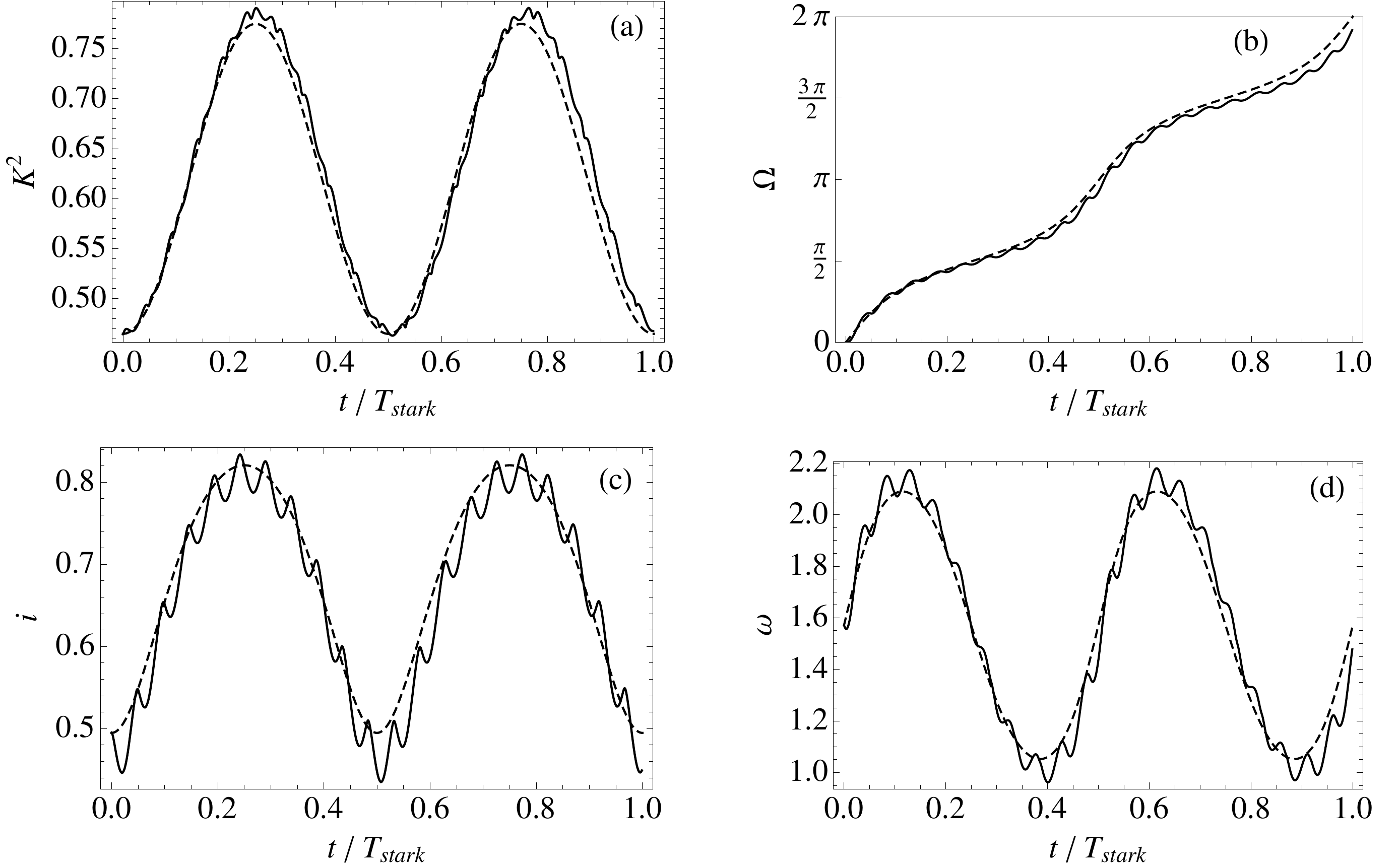}
  \caption{Plots of (a)
  $K^2(t)$, (b) $\Omega(t)$, (c) $i(t)$, and (d) $\omega(t)$ for an orbit
  with $K_z = 0.6$, $\Delta \tilde H = 0.34$, and $\alpha = 0.04$.
  Dashed lines show results from orbit-averaged theory
  (equations [\ref{eq:omegadot}] - [\ref{eq:OofK}]), and the solid lines
  show results from numerical integrations.}
\label{3Dstarkfig}
\end{figure}

Figure \ref{3Dstarkfig} compares the analytic expressions for 
$K^2(t)=1-e^2(t)$, $i(t)$, $\omega(t)$, and $\Omega(t)$ with numerical
integrations. Good agreement between 
the two approaches can be seen if one disregards the small-amplitude
oscillations
of orbital elements on a time scale $T_K$ which are not captured by 
our orbit-averaged theory. 

We have assumed in \S\S \ref{subsect:Stark_planar} and
\ref{subsect:Stark_3D} that the Stark vector is constant, but in
reality it can vary in both magnitude and direction due to effects
such as a changing relative velocity between the Solar System and the ISM, a
variation in the particle's charge as a function of distance from the
Sun, small scale magnetic turbulence in the ISM, etc. We study these
effects in Appendix
\ref{starkvariations} and conclude that most likely they add details to, but
do not significantly change the general picture presented in \S\S
\ref{subsect:Stark_planar} and \ref{subsect:Stark_3D}.

\subsection{The general Stark problem: particle ejection}
\label{particleejection}

When $\alpha\sim 1$ our perturbative approach used in
\S\ref{subsect:Stark_planar}-\S\ref{subsect:Stark_3D}
breaks down. In fact, for $\alpha\gtrsim 1$ one expects 
the particle to become unbound since the Stark force 
becomes larger than $F_g$. We now perform a detailed treatment of
particle ejection with the goal of determining the value of $\alpha$
that yields ejection for different initial conditions.

\begin{figure} [h] 
  \centering
  \includegraphics[width=.5\textwidth]{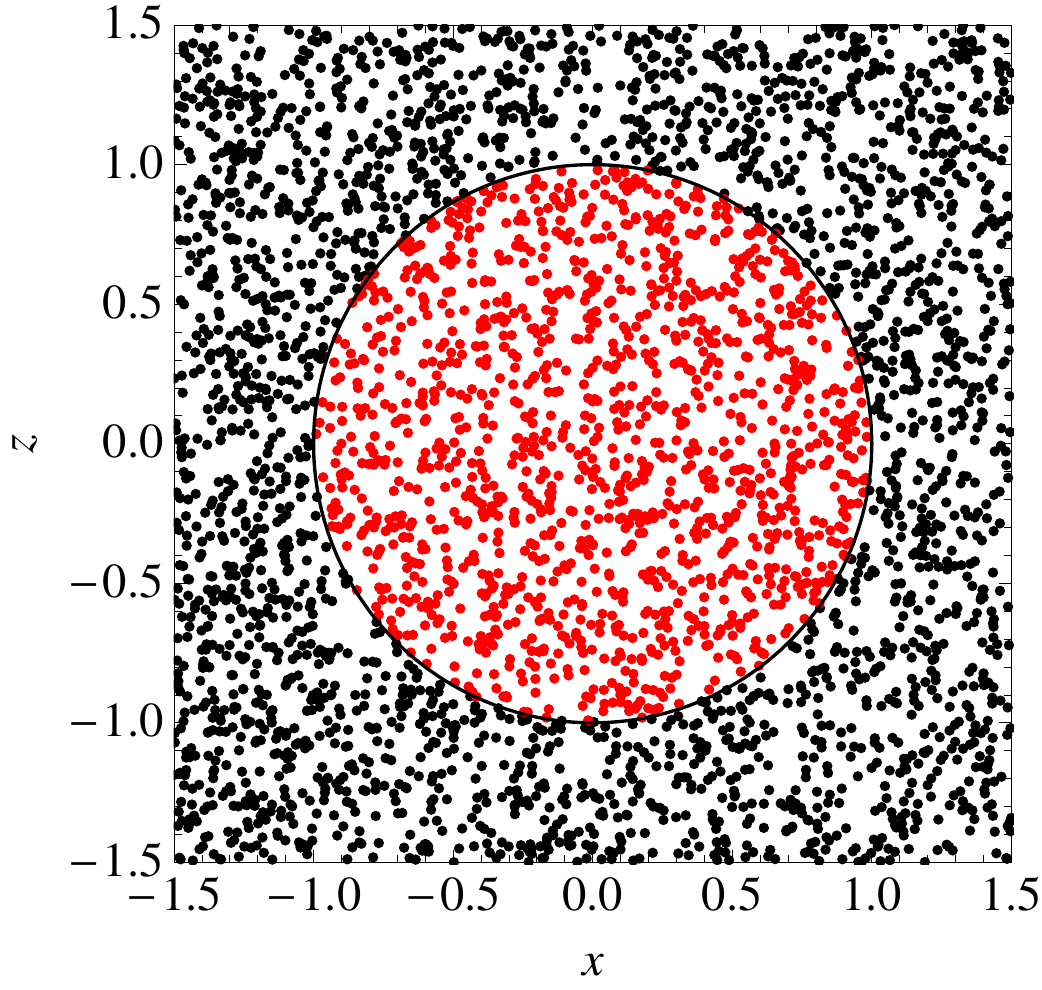}
  \caption{Illustration of the ejection criterion for particles
  initialized at rest, showing the outcome of the numerical
  integration of an orbit (black - ejection, red - orbits bound after
  an integration time of 10 $T_K$). The Stark force is directed along
  the z-axis and because of rotational symmetry we only need to
  consider
  particles initialized in the x-z plane. The black circle
  shows the radius of the sphere beyond which orbits are unbound as
  given by the ejection criterion (equation
  [\ref{eq:ejsimple}]), and the units on the axes are scaled so that
  the sphere has a radius of one. The numerical integrations agree
  with equation (\ref{eq:ejsimple}), since only those particles
  initialized outside of a radius of one are ejected within 10
  $T_K$ (most unbound particles are ejected 
  within 1 $T_K$, so virtually all unbound particles should be ejected
  after 10 $T_K$.). }
  \label{ejectionfig}
\end{figure}

\citet{Kirchgraber}, \citet{Dankowicz}, and \citet{Banks} have previously
made attempts to study ejection in the Stark problem but their 
results are not easy to interpret. In Appendix \ref{app:ejection}
we show that using the approach of \citet{Kirchgraber} it is still possible 
to derive a set of simple analytical criteria for testing whether 
a particle with a given initial position ${\bf r}^0$ and velocity
${\bf v}^0$ is bound or unbound (equations 
[\ref{eq:pot}]-[\ref{eq:bound}]). In finding these criteria we did
not use the rather complicated solutions for particle 
motion obtained in \citet{Kirchgraber}, but instead resorted to 
general energetic arguments applied to the particle Hamiltonian expressed 
in Kustaanheimo-Stiefel variables (Kustaanheimo \& Stiefel 1965). 

The Stark force introduces a preferred direction into the 
otherwise isotropic Keplerian problem. Thus, one expects that 
in general the ejection criterion would depend not only on the 
initial particle distance from the BSS but also on the
direction of ${\bf r}^0$ and on ${\bf v}^0$. Quite interestingly,
we find in Appendix \ref{app:ejection} that in the case 
of dust grains starting at rest with respect to the BSS (${\bf v}^0=0$) 
the ejection criterion is {\it independent} of the direction of 
${\bf r}^0$ and simply states that all particles with 
\ba
\alpha > 0.25,~~~~~|{\bf v}^0|=0,
\label{eq:ejsimple}
\ea
are going to be ejected from the system. Figure \ref{ejectionfig}
illustrates this simple ejection criterion 
computationally: orbits of a number of particles with randomly chosen
${\bf r}^0$ and $|{\bf v}^0|=0$ have been integrated for ten
orbital timescales each
to determine whether they are bound or not; the condition 
(\ref{eq:ejsimple}) clearly works quite well. 

\begin{figure}[!p] 
  \centering
  \begin{tabular}{cc}
  \includegraphics[width=.4\textwidth]{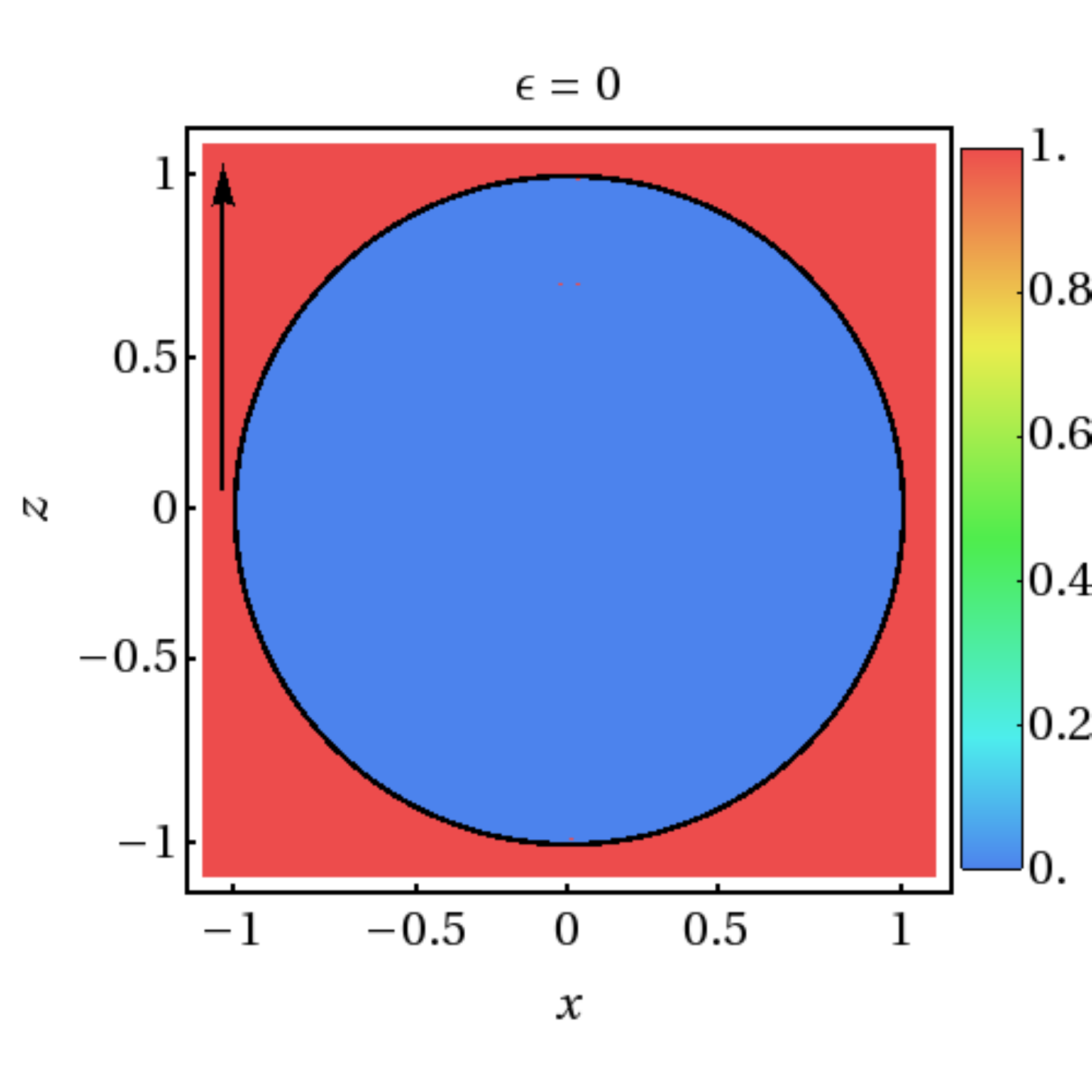} &
  \includegraphics[width=.4\textwidth]{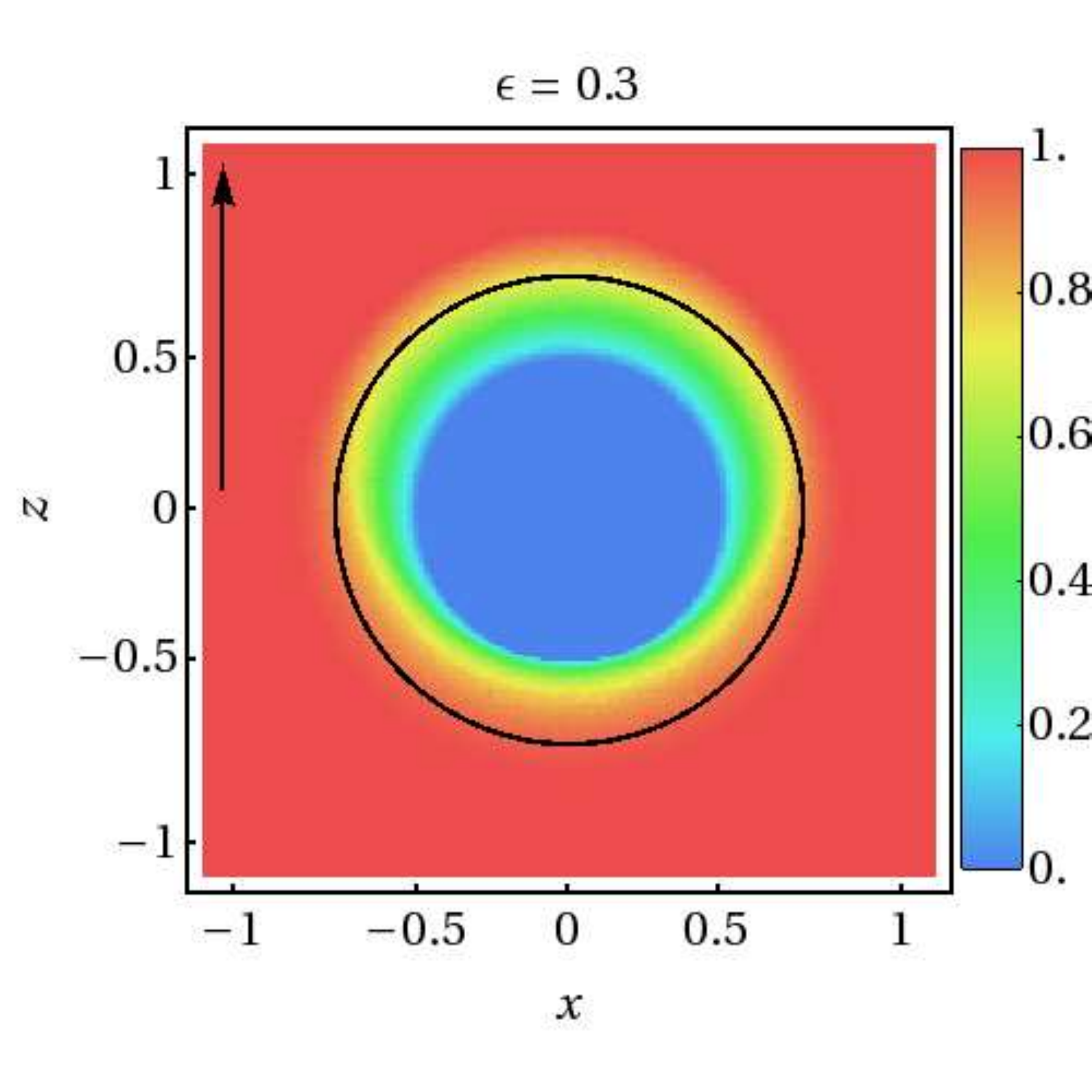} \\
  \includegraphics[width=.4\textwidth]{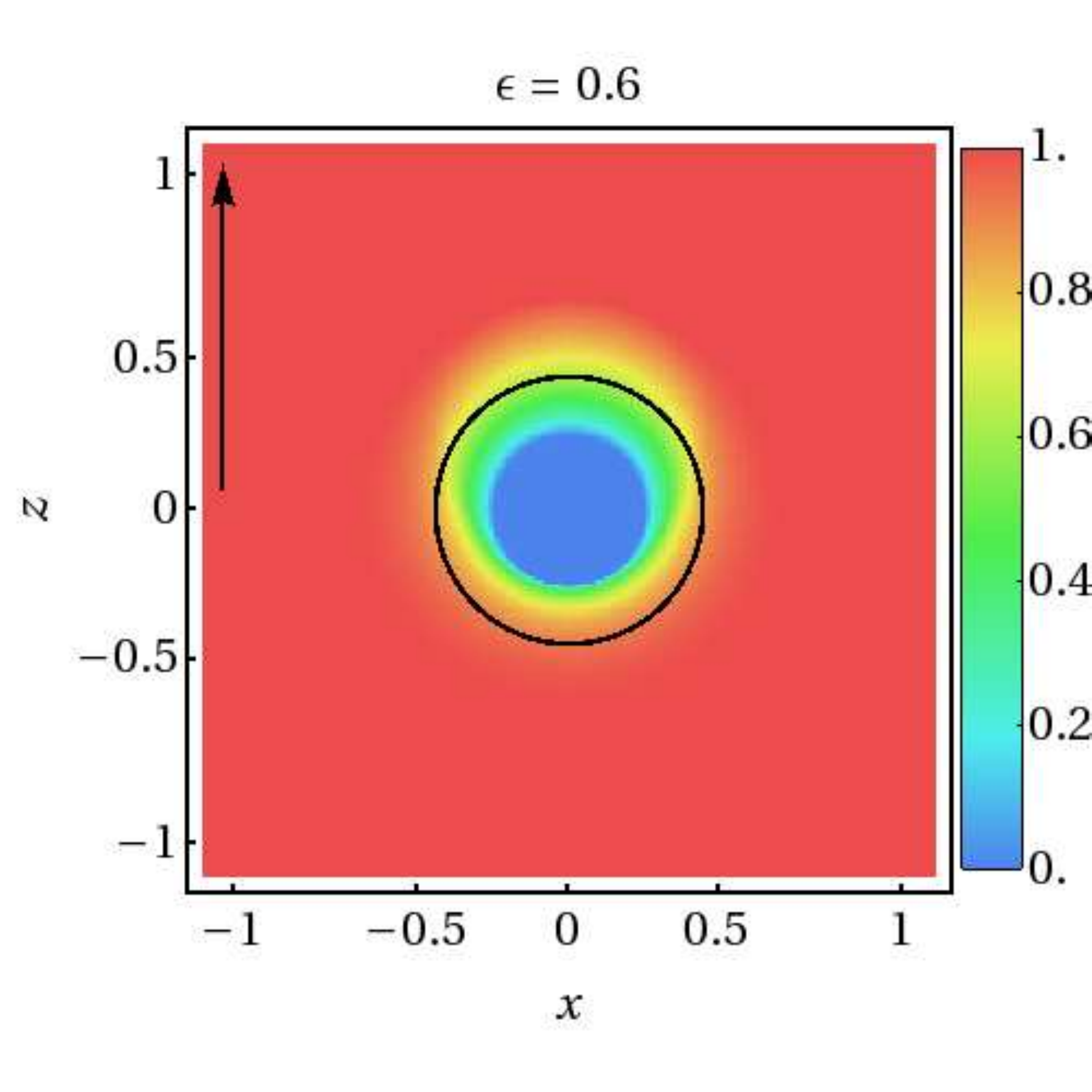} &
  \includegraphics[width=.4\textwidth]{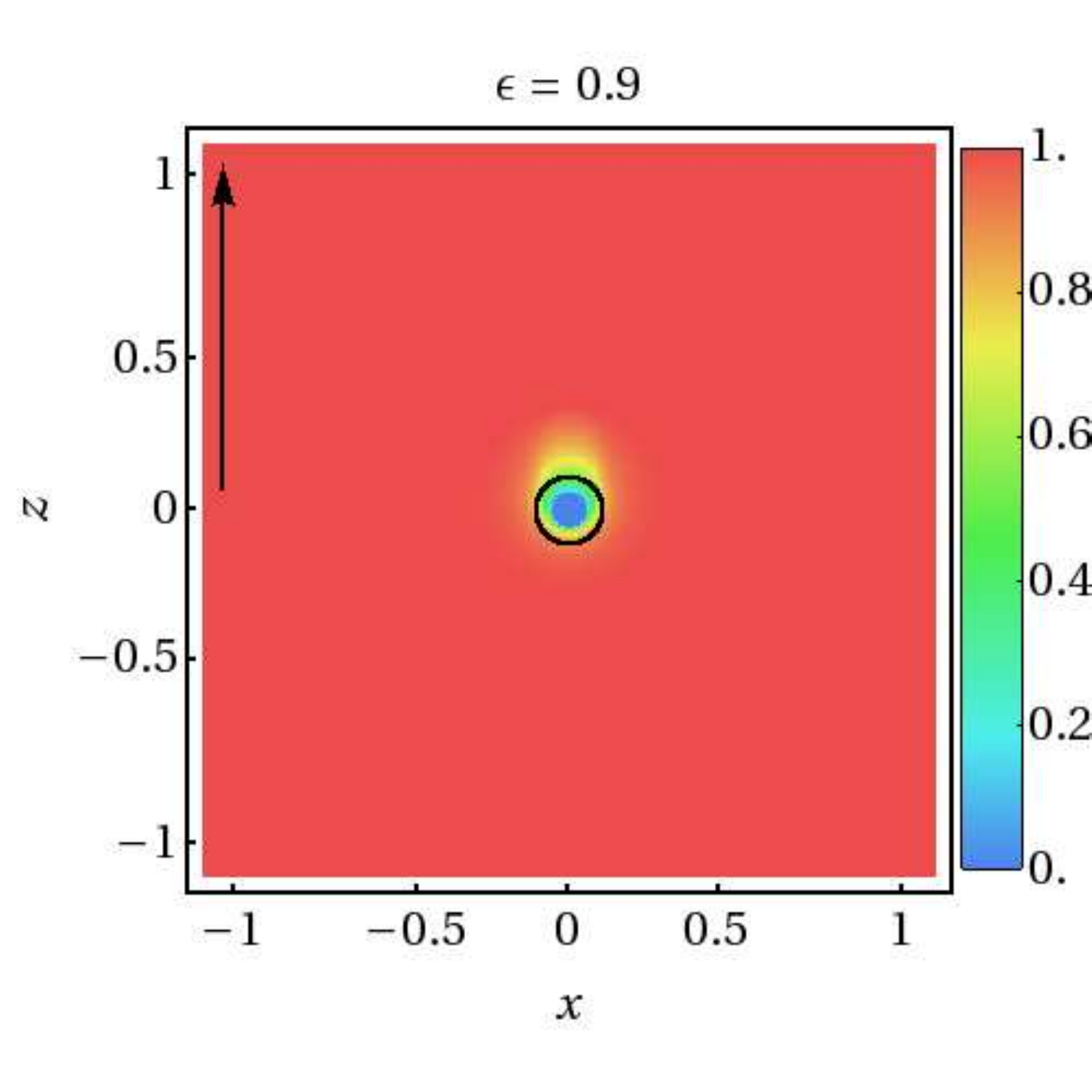} \\
  \end{tabular}
  \caption{Ejection in the case of
  non-zero initial velocity. The Stark vector points in the $+z$
  direction, as indicated by the arrows, so the problem is rotationally symmetric about the z-axis,
  and we confine ourselves to the x-z plane. Each panel shows the fraction of ejected particles as a
  function of starting position for a given value of $\epsilon$
  (defined in the text). Red corresponds
  to all particles initialized at a given location being ejected and blue corresponds to all particles
  being bound after 10 $T_K$. The scale of the $x$ and $z$ axes is the same in each
  plot and is normalized by the value of the semi-major axis at which
  $\alpha=0.25$ in the zero velocity case ($\epsilon =0$).
  The black circle shows the $\alpha=0.25$ contour for the value of $\epsilon$
  with which particles are initialized (it contracts inwards
  with increasing $\epsilon$ because of the relation $a^0 = r^0/2(1-\epsilon)$).}
  \label{KEcontoursfig}
\end{figure}

To study the case of a nonzero initial
velocity, we performed Monte Carlo simulations where we initialized
particles at a given radius to have a random orientation of ${\bf v}^0$
and a kinetic energy which is a fixed fraction $\epsilon$ of their 
gravitational potential energy: $\left(v^0\right)^2/2 = \epsilon GM_\odot/r^0$. 
The initial semi-major axis $a^0$ can then be expressed in terms of 
the initial radius as $a^0 = r^0/2(1-\epsilon)$, which
allows the calculation of $\alpha$ in each case.
The case of zero initial velocity corresponds to $\epsilon=0$. 
In all non-zero initial velocity 
cases simulated here we found the general analytical ejection 
criteria (\ref{eq:pot})-(\ref{eq:bound}) to correctly predict whether
particles are bound to the Solar System.

\begin{figure}[h] 
  \centering
  \includegraphics[width=.5 \textwidth]{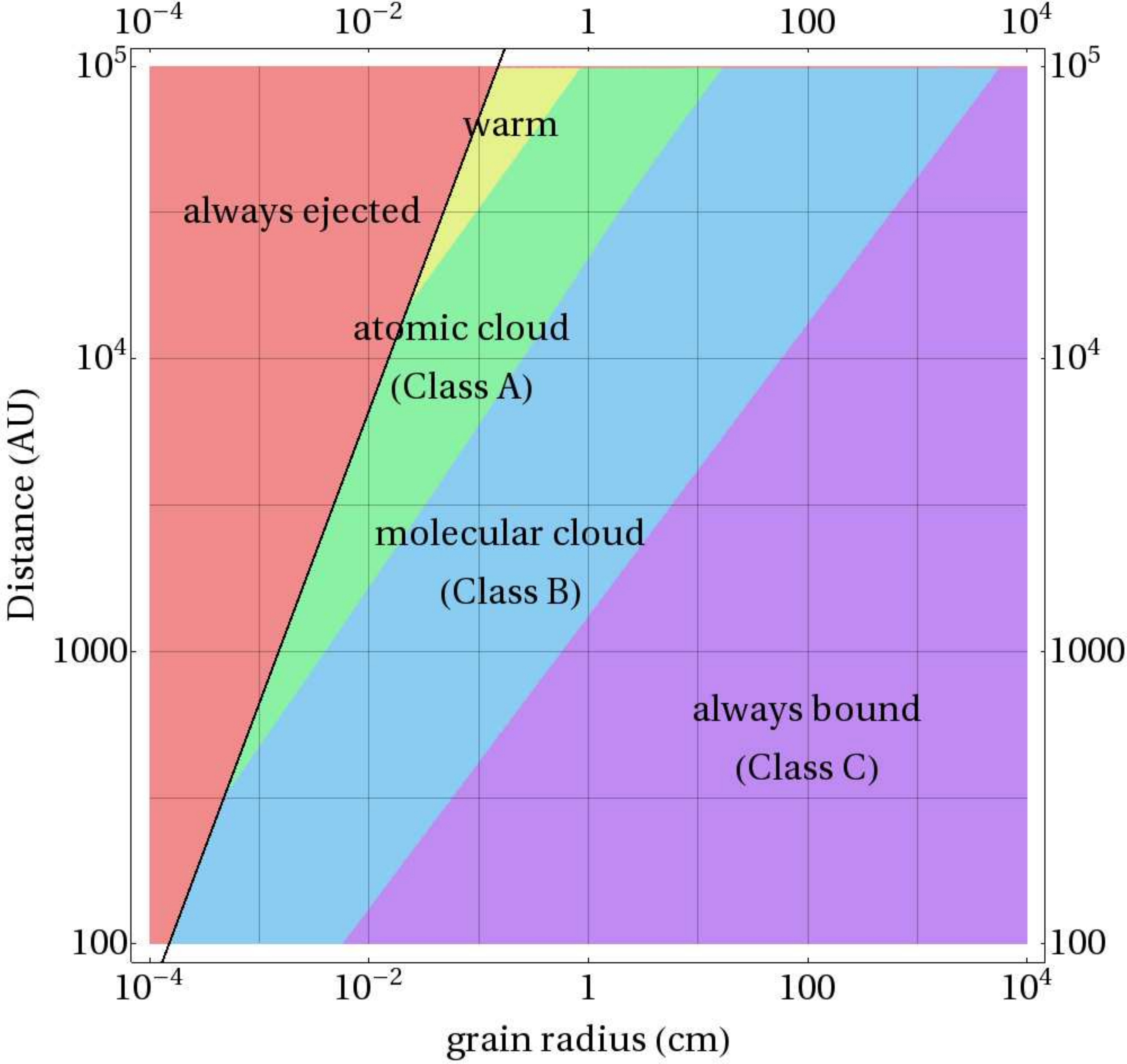}
  \caption{Map of the ejection regimes in particle radius -
  orbital distance space. In the purple region, particles
  are bound regardless of the ISM phase (Class C
  particles, \S\ref{casethree}). In the adjacent light blue
  region, particles will be ejected when passing through a
  molecular cloud, but will be bound in all other phases (Class B
  particles, \S\ref{casetwo}).
  In the green region, particles will be ejected in atomic
  clouds and molecular clouds, but will be bound otherwise (Class A
  particles, \S\ref{caseone}). In the
  yellow region particles are only bound in the coronal phase (also Class A
  particles, \S\ref{caseone}). In the
  red region, particles are unbound in all ISM phases, and the
  boundary of this region is given by equation (\ref{rgminE}), which is shown as a black line
  on the plot.}
  \label{ISMejection}
\end{figure}

Figure \ref{KEcontoursfig} shows the effect of varying $\epsilon$ 
on the value of $\alpha$ required for ejection. In the case of  
$|{\bf v}^0| \ne 0$, each initial particle position 
is characterized by some {\it probability} of ejection since whether
a body is ejected or not depends on the orientation of ${\bf v}^0$,
and we display these probabilities with color maps. Most notably, 
Figure \ref{KEcontoursfig} shows that for $\epsilon > 0$ there 
exists an outer boundary beyond which
all particles are ejected regardless of their initial velocity, an
inner boundary inside of which all orbits are bound for $\epsilon <
1$, and a region in between in which orbits can be bound or unbound
depending on the orientation of ${\bf v}^0$. As ${\bf v}^0$ increases and
$\epsilon$ becomes larger, the inner and outer ejection boundaries
become more elongated along the direction of ${\bf S}$, and ejection 
can occur for certain orbits even if $\alpha < 0.25$, whereas some orbits 
can remain bound even if $\alpha > 0.25$. However, even for $\epsilon$
as high as $0.9$ the simple ejection criterion (equation [\ref{eq:ejsimple}]) still
provides  a 
good estimate of the radius at which particles become unbound to
within a factor of two. 

In Figure \ref{ISMejection} we use the ejection criterion
from equation (\ref{eq:ejsimple}) to determine whether particles are bound or unbound
at different separations from the BSS in different ISM phases. It clearly
demonstrates that independent of the inflowing ISM phase, particles as large as 
$r_g=150\,\mu$m should be ejected by the electromagnetic force at $a=10^4$ AU, while at $a=10^3$ 
AU only particles with $r_g \lesssim 15 \,\mu$m are unbound (equation
[\ref{rgminE}]). Bigger grains can also be 
removed in passages through the denser phases of the ISM. In particular,
passages through molecular clouds would eject bodies as large as
$1$ m in size at $10^4$ AU due to the total drag force, in agreement
with \citet{Sternerosion}.

Very small particles for which $\alpha\gg 1$ should be rapidly coupled
to the ISM flow by the non-gravitational forces. We introduce the {\it coupling particle size}
$r_{g,{\rm coupl}}$  
as the size for which the coupling distance is equal to the initial 
particle semi-major axis, i.e. $d_{\rm coupl}(r_{g,{\rm
    coupl}})=a$. Particles with $r_g \lesssim r_{g,\text{coupl}}$
rapidly attain a velocity $\sim v_w$ in the solar frame. This makes it
necessary to consider the Stark force as a function of velocity and
take the magnetic force into consideration, since the relation $v \sim
v_K \ll v_w$ is no longer valid.
The dependence of $r_{g,{\rm coupl}}$ on $a$ is shown in Figure 
\ref{radiusradius} by a dashed line. 
It is clear that efficient coupling to the ISM
on scales of order the initial size of the system is possible only 
for particles with $r_g$ considerably smaller than the size at which
ejection occurs. Particles with $r_g\gtrsim r_{g,{\rm coupl}}$ 
passing close to the Inner Solar System should have speeds considerably
different from (smaller than) $v_w$.


\section{Orbital Decay via the Differential Drag}
\label{sect:orbdecay}


We now consider the effect  of the 
differential drag force $\Delta {\bf F}_\text{drag}$ on the motion of
dust particles. $\Delta {\bf F}_\text{drag}$ explicitly 
depends on the Keplerian velocity of the grain, and plugging the
expression for $\Delta {\bf F}_\text{drag}$ from equation
(\ref{masterdrag}) into the equations for the evolution of the
osculating elements \citep{Burns,MD}, one obtains the following
equation for the orbit-averaged rate of change of the semi-major
axis
\begin{eqnarray}
\label{axisdecay}
\left \langle \frac{da}{dt} \right \rangle  = -\frac{2a}{m_g v_w}\left[F(v_w)
(1-\beta) +  \frac{d F}{d v} \Big |_{v=v_w} v_w \beta \right],
\label{eq:dadtdifdrag}
\ea
where we have defined
\ba
\beta \equiv e^{-2}\left[\left(1-\sqrt{1-e^2}\right) 
\left(\frac{{\bf v}_w \cdot \hat{\bf a}}{v_w}\right)^2 + \left(e^2 - 1 
+ \sqrt{1-e^2}\right) \left(\frac{{\bf v}_w \cdot
  \hat{\bf b}}{v_w} \right)^2\right],
\label{eq:betadef}
\end{eqnarray}
and ${\bf \hat{a}}$, ${\bf \hat{b}}$,
and ${\bf \hat{c}}$ are orthonormal unit vectors with ${\bf \hat{a}}$
pointing towards the pericenter, ${\bf \hat{c}}$ aligned with the 
particle's angular momentum vector, and ${\bf \hat{b}} = {\bf \hat{c}} \times
{\bf \hat{a}}$ \citep{Burns}. Equation (\ref{eq:dadtdifdrag}) agrees with the results of P\'astor 
\etal 2010 who obtained it for the special case of a drag force that depends quadratically
on the velocity. 

The parameter $\beta$ has the property
that $0 \leq \beta \leq 1$ for any $e < 1$. Although $\beta$ 
varies on timescales of order $T_\text{stark}$ because of the secular 
changes in eccentricity and orientation of the orbit, we can average
the right-hand side of equation (\ref{axisdecay}) over a Stark orbit to obtain
\begin{eqnarray}
\label{eq:simpledecay}
  \left \langle a(t) \right \rangle &=& a_0 e^{- \gamma t} \\
  \label{decayconstant}
  \gamma &\equiv& \frac{2}{m_g v_w}\left[
  F(v_w)(1-\langle \beta \rangle_S) + \frac{d F}{d
  v} \Big |_{v=v_w} v_w \langle \beta \rangle_S \right], 
\end{eqnarray}
where angle brackets with a subscript $S$ denote time-averaging over a
Stark orbit, angle brackets without a subscript denote
time-averaging over a Keplerian orbit, and $a_0$ is the initial value
of the semi-major axis. The validity of averaging over
a Stark orbit in equation (\ref{eq:simpledecay}) hinges upon
$\gamma^{-1} \gg T_\text{stark}$, since this is the condition for the
orbit to be only slightly modified by the differential drag over
a timescale $T_\text{stark}$. This condition is satisfied in practice,
since according to equations (\ref{starkperiod}) and
(\ref{decayconstant})
\begin{equation}
\gamma T_\text{stark} \sim \frac{v_K F_\text{drag}}{v_w S} \ll 1,
\end{equation} 
where $S \gtrsim F_\text{drag}$ ($S > F_\text{drag}$ if, for example,
the electric force dominates the total drag force). Thus, as long as $v_K/v_w
\ll 1$, the semi-major axis evolves exponentially with an instantaneous
decay constant given by $\gamma$.

An immediate
consequence of equation (\ref{decayconstant}) is that the semi-major axis
can either grow or decay in time depending on the value of $\beta$ and
the sign of $dF/dv$. For gas drag, $dF/dv> 0$ for any $v$, so
$\gamma > 0$, and the orbit always decays with time. On the 
other hand, for pure Coulomb drag $dF/dv<0$ if the Mach number 
of the ISM flow with respect to the sound speed of the ionized 
component $s \gtrsim 0.968$ (equations 
[\ref{qeq}]-[\ref{eq:Lambda}]). Thus, depending on the value of 
$\beta$, the semi-major axis can either grow or decay 
with time. 

\begin{figure}[h] 
  \centering
  \includegraphics[width=.5 \textwidth]{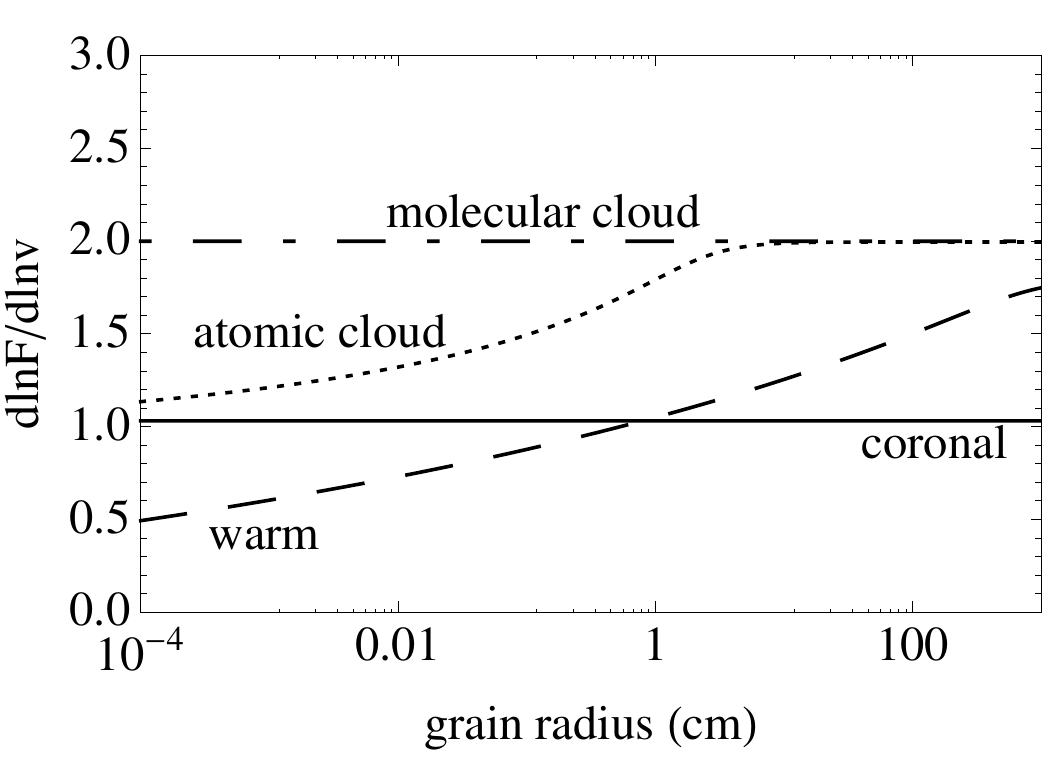}
  \caption{Plot of $d \ln F/d\ln v |_{v=v_w} $ as a function of
  grain size for various ISM phases, where $F$ is the sum of Coulomb
  and gas drag forces. Because $d \ln F/d\ln v 
  |_{v=v_w} > 0$ over the whole range of grain sizes for each
  phase of the ISM, the differential
  drag always causes the semi-major axes of particle orbits to decay, 
  see equation (\ref{axisdecay}).} 
  \label{dFdvfig}
\end{figure}

An interesting question is whether differential drag can 
cause the {\it expansion} of particle orbits under realistic ISM 
conditions. According to equation (\ref{decayconstant}) this amounts 
to a calculation of $dF/dv|_{v=v_w}$ where $F$ is the magnitude of the
force
due to both gas and Coulomb drag combined under actual ISM
conditions. The results of such a calculations are presented in 
Figure \ref{dFdvfig} and clearly show that for any grain size in 
any of the ISM phases $d F/dv|_{v=v_w} > 0$, meaning that 
a particle's semi-major axis can only {\it decay} ($\gamma > 0$)
under the action of the differential drag force.

\section{Application to Grains in the Outer Solar System}
\label{dgdapplication}

Having explored the forces that affect the dynamics of dust
grains in the OSS, we are now in a position to describe
their long-term evolution after they have been created in collisions
of larger objects (e.g. comets). Our discussion will predominantly 
focus on two important 
questions. First, what is the survival time $T_s$ of a dust particle 
of a given size located at a given separation from the BSS against 
ejection by the ISM flow. Second, how does the semi-major axis of 
a particle change as a result of differential
drag during its lifetime in the OSS? 

In addressing these questions we must realize that through 
its 4.5 Gyr history, the Solar System has sampled different phases of the 
ISM, which must have subjected dust grains in the OSS to broadly 
varying ambient conditions. As a result, grains which are only 
weakly affected by the ISM flow when the Solar System passes through one 
phase (e.g. the coronal gas) may become fully coupled to this 
flow in another phase (e.g. a molecular cloud), see Figure 
\ref{radiusradius}. Therefore there are several possible classes into
which dust grains can be separated from the point of view of their 
long-term dynamics: 
\begin{itemize}

\item Class A covers particles which can survive ejection in
the warm and coronal phases of the ISM, but which will be 
ejected in a passage through an atomic or molecular cloud.

\item Class B involves particles which are large enough to survive 
ejection in atomic clouds, but which would be ejected in passages 
through molecular clouds.

\item Class C covers particles which are large enough to survive
ejection in a molecular cloud. Thus these particles can survive
ejection in any phase of the ISM. 

\end{itemize}
The assignment of a given dust grain to a particular class depends 
not only on the particle size but also on the particle distance 
from the BSS. Figure \ref{ISMejection} shows the map of 
different Classes in the $r_g-a$ space. Note that
particles satisfying the condition $r_g < r_{g,\text{min}}$ with
$r_{g,\text{min}}$ defined in equation (\ref{rgminE}) do not exhibit
long-term dynamics on the OSS as they get rapidly
ejected by the electric force even in the coronal phase of the ISM. For that
reason we do not introduce a separate class for these small grains.

\subsection{Class A}
\label{caseone}

\citet{Sternerosion} quotes $40$ Myr for the average 
time between passages through 
atomic clouds. This is roughly survival time, $T_s$, of Class A
particles since these are ejected in passages through atomic or
  molecular clouds, but passages through molecular clouds happen on
  much longer, $\sim 400 \Myr$ \citep{Sternerosion}, timescales.

We now study the decay in semi-major axis caused by the differential
drag in the coronal and warm phases of the ISM in between
ejection events. We consider particles at $3000$ AU with $\rho =
1$ g cm$^{-3}$ and $\alpha =0.05$, which corresponds to a grain radius 
$r_g\approx 100 \mm$ for both the coronal and warm
phases. The reason the grain radii are the same in different phases 
is because the electric force is the dominant perturbation in both 
cases, and we are assuming the particles are charged to the same potential
of $1$ V. Our choice of $\alpha$ is close to the threshold indicated 
in the ejection criterion (\ref{eq:ejsimple}) and is motivated by the 
fact that differential drag is most important for the smallest 
particles.

Equation (\ref{eq:simpledecay}) predicts roughly exponential 
decay of the particle orbit on some characteristic time scale 
$t_\text{decay}=\gamma^{-1}$. To evaluate upper and lower limits for 
$t_\text{decay}$, we calculate $\gamma^{-1}$ in equation 
(\ref{decayconstant}) for $\beta = 0$ and $\beta = 1$. We 
can immediately rule out differential drag playing an 
important role in the coronal phase, since 
$2.8 \Gyr < t_\text{decay} < 3.0 \Gyr$ there, which is much 
longer than the $40$ Myr interval between ejection events. On the 
other hand, a similar calculation for the warm phase yields 
$27 \Myr <t_{A,\text{decay}} < 38 \Myr$. Since this is smaller than 
$T_s\sim 40 \Myr$, the semi-major axes of some Class A particles 
can decay by a factor of a few between ejection events. 

To verify our estimate for $t_\text{decay}$ in the warm phase, 
we perform Monte Carlo simulations of dust grain orbital evolution.
We initialize particles to have a semi-major axis of $3000 \AU$, 
  corresponding to the inner edge of the Oort Cloud where most of 
the dust should be concentrated. Without good 
knowledge of what the distribution of
  the other orbital elements should be, we choose the 
eccentricities uniformly on the interval $(0,0.99)$, and randomly
orient the orbit in space. The perturbations included in the simulation on
top of the gravitational attraction to the BSS are
the electric force, gas drag, and Coulomb drag; other perturbations
forces are unimportant for particles of this size at this semimajor axis.

\begin{figure}[h] 
  \centering
  \subfigure{\label{fig:dfa}\includegraphics[width=0.3\textwidth]{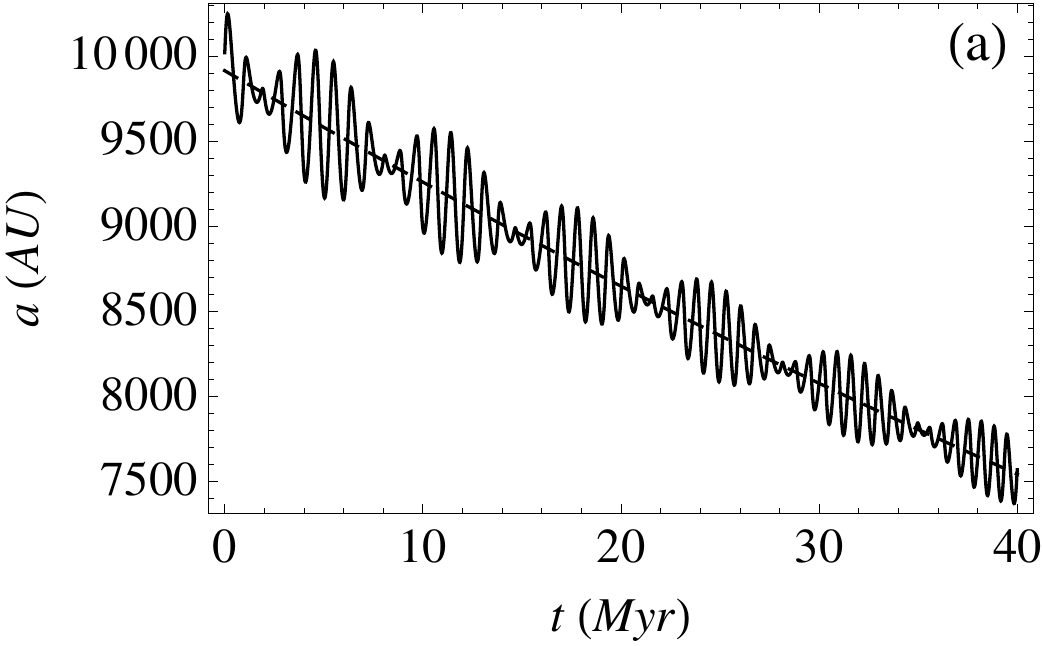}}
  \subfigure{\label{fig:dfb}\includegraphics[width=0.3\textwidth]{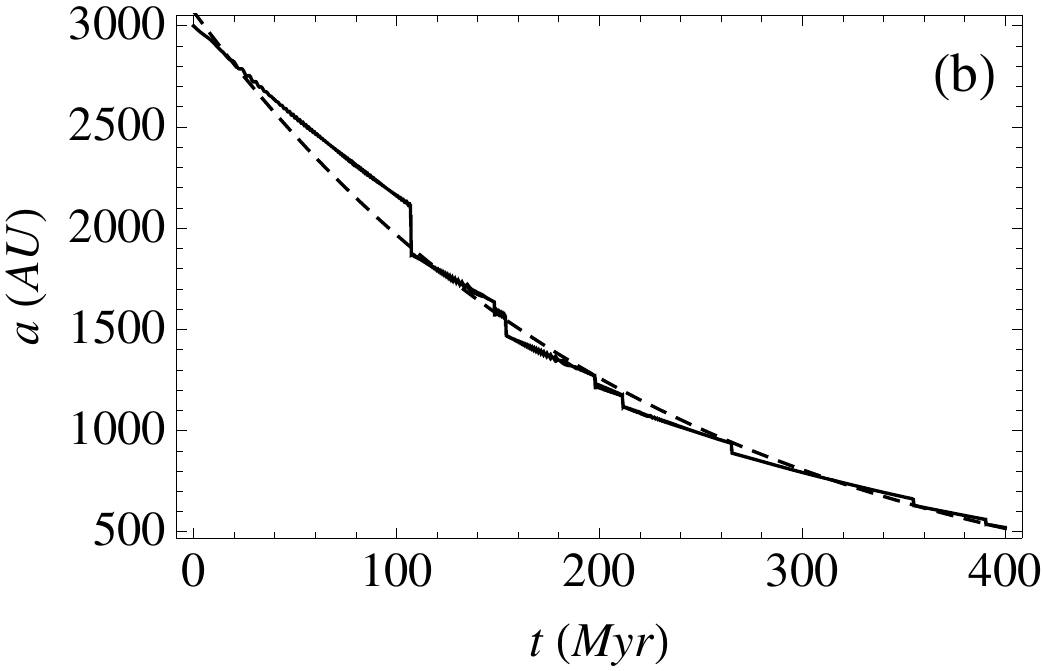}}
  \subfigure{\label{fig:dfc}\includegraphics[width=0.3\textwidth]{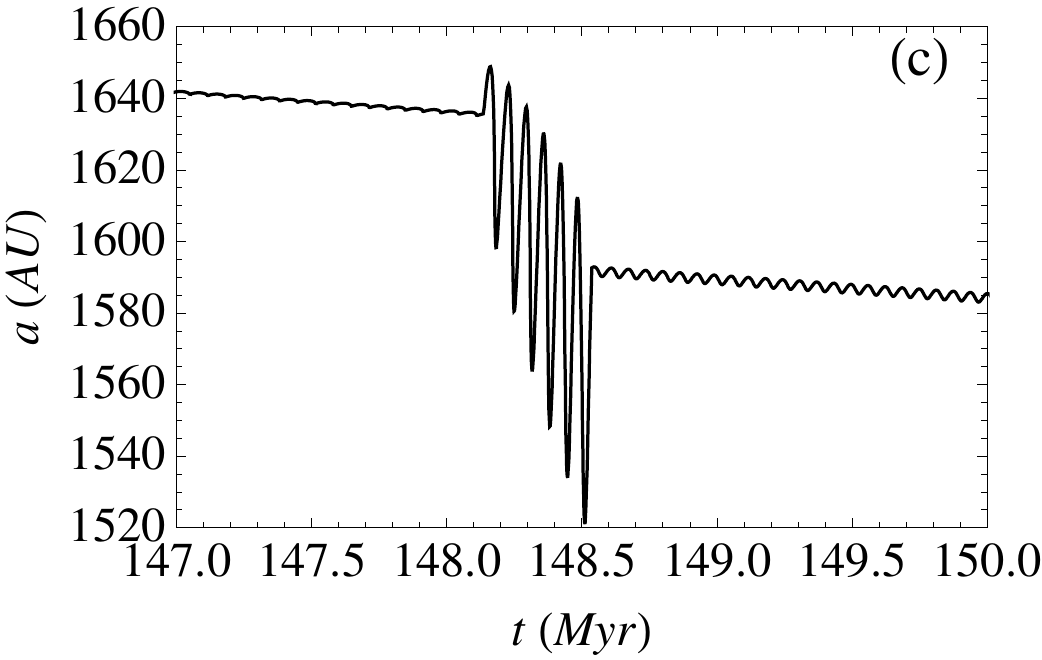}}
  \caption{(a) The decay of a particle with an
  initial semi-major axis of $10^4 \AU$ and $\alpha = 0.05$ over $40
  \Myr$ in the warm phase. The rapid oscillations are on an orbital
  timescale and the long oscillations are on a Stark period
  timescale. The
  dashed line is the best fit exponential to the decay.
  (b) The decay of a Class B particle. The jumps in semi-major
  axis occur when the particle is passing through an atomic cloud. The
  dashed line is again the best fit exponential to the decay.
  (c) Zoomed in view of one of the passages through an
  atomic cloud from panel (b).}
  \label{decayfig}
\end{figure}

We simulate $10^4$ orbits 
for $40$ Myr assuming particle parameters adopted for Class A (see above). For
each run we compute the average $\gamma$ over the length of the
simulation by finding a best exponential fit to the dependence 
of the semi-major axis on time. A typical example for 
the evolution of the semi-major 
axis of a particle having $\alpha = 0.05$, but with a
starting semi-major axis of $10^4 \AU$ rather than $3000 \AU$ is shown
in Figure \ref{decayfig}a. The reason for choosing $10^4 \AU$ for the
starting semi-major axis is to
highlight the rapid oscillations happening on an orbital
timescale, the long oscillations occurring on a Stark period
timescale, and the steady decay due to differential drag. The 
resulting distribution of decay times, $t_\text{decay}=\gamma^{-1}$, is
plotted in Figure \ref{decayhist}a. We calculate from this distribution 
that the mean decay time for Class A grains is $\langle t_\text{decay}
\rangle = 30.4$ Myr with a 
standard deviation of $1.1$ Myr. This agrees nicely with the analytical 
limits given above. 

Furthermore, we see from equation
 (\ref{decayconstant}) that the {\it limits} on $t_\text{decay}$ do not
  depend on the semimajor axis. Thus, as long as a grain is not 
eroded or destroyed in
a collision, and the limits on $t_\text{decay}$ are tight, $\gamma$
will be approximately constant for a grain of a given size as it
decays. Of course, smaller and smaller particles become bound as the
semimajor axis decreases, so for $\alpha = 0.05$ ($r_g \approx 10 \mm$)
at $300$ AU in the
warm phase, the limits on $t_\text{decay}$ are a mere $2.6 \Myr < 
t_\text{decay} < 4.3 \Myr$, whereas at $10^4$ AU for $\alpha = 0.05$
  ($r_g \approx 500 \mm$) they are $150 \Myr < 
t_\text{decay} < 180 \Myr$, substantially longer than the timescale
between passages through atomic clouds.

The orbital decay of Class A particles should lead to their preferential 
concentration at small radii, which may promote fragmentation of 
these grains in mutual collisions and the creation of even smaller debris 
particles closer to the Sun. The implications of this effect are 
discussed in more detail in \S\ref{satellites}.

\subsection{Class B}
\label{casetwo}

Class B comprises particles which are large enough
to survive ejection in atomic clouds, but not in molecular
clouds. The time between the Solar System's passages through molecular clouds is
about $400 \Myr$ \citep{Sternerosion}, so this is
the survival time for Class B dust grains.

We take the parameters of a typical Class B particle to be $\rho = 1
\gcm$ and $r_g\approx 0.1 \cm$, since such a particle would have 
$\alpha = 0.05$ in an atomic cloud at $a=3000 \AU$. 
Prior to ejection in a molecular cloud, the particle will spend a
fraction $f_\text{atomic}$ of its lifetime in atomic
clouds, where $f_\text{atomic}\sim 0.01$ is the filling fraction of these clouds,
and a fraction $1-f_\text{atomic}$ of its lifetime in the warm and coronal phases combined,
see Table \ref{ISMphases}. For simplicity, we assume that the
particle spends its time in either the warm
phase or an atomic cloud before it is ejected\footnote{Assuming the particle samples the
  coronal phase for a fraction of time $f_\text{coronal} \sim 0.5$
  would raise $t_\text{decay}$ by approximately a factor of 2
  because the decay in the coronal phase is negligible.}.
Then, the characteristic decay time 
is $t_{\text{decay}} =[(\gamma_{\text{WISM}}(1-f_\text{atomic}) +
\gamma_\text{atomic}f_\text{atomic})]^{-1}$, where $\gamma_{\text{WISM}}$ is the 
value of the decay constant $\gamma$ in the warm phase and 
$\gamma_\text{atomic}$ is $\gamma$ in atomic clouds. Even though 
$f_\text{atomic} \ll 1$ the contribution to the decay from atomic clouds is
still important since the total drag in them is much larger than in the 
more dilute phases of the ISM. Setting as in \S\ref{caseone} $\beta =
0$ and $\beta = 1$ in equation (\ref{decayconstant}) to get upper and
lower bounds on the decay time we have $ 205 \Myr <
t_{B,\text{decay}} < 255 \Myr$. This is shorter than the $T_s = 400$ Myr time 
between passages through a molecular cloud, so the orbits of these 
particles can also decay appreciably before ejection occurs.   

We checked our analytical estimates by simulating $10^4$ particle 
orbits for $400 \Myr$ initialized in the same way as in \S\ref{caseone} (with
Class B particle parameters adopted here).  
We simulate the effect of passages through atomic clouds as stochastic
events that happen on average every $50$ Myr, so that there are on average 
8 passages through atomic clouds per simulation, and the total time spent
in atomic clouds per simulation is on average $4$ Myr, consistent with 
$f_v = 0.01$. A typical example of the orbital evolution of a Class B
dust particle along with the best exponential fit is shown in Figures
\ref{decayfig}b,c. 

\begin{figure}[h] 
  \centering
  \subfigure{\label{dha}\includegraphics[width=0.3\textwidth]{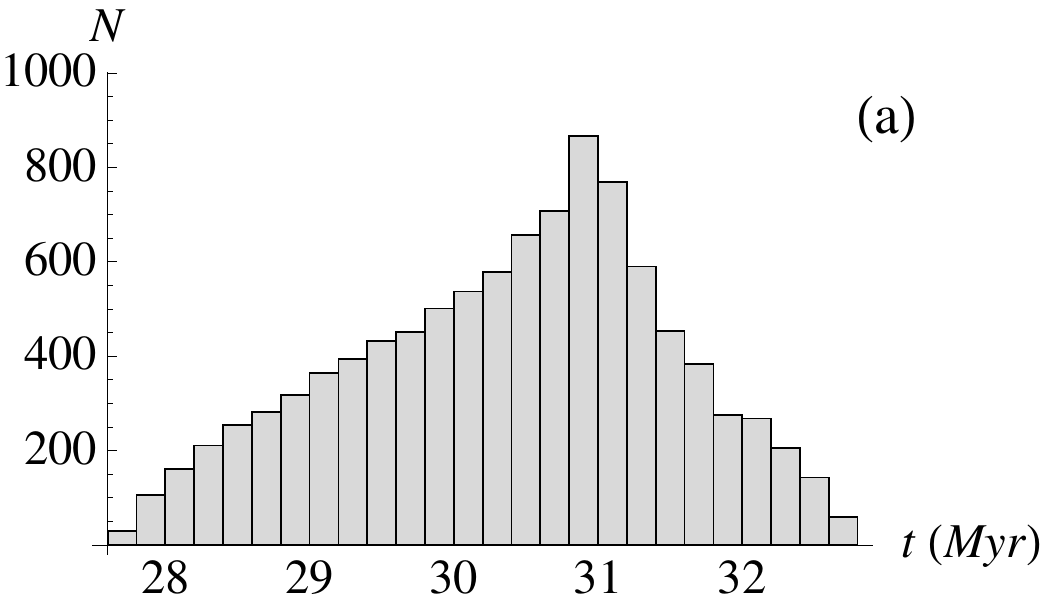}}                
  \subfigure{\label{dhb}\includegraphics[width=0.3\textwidth]{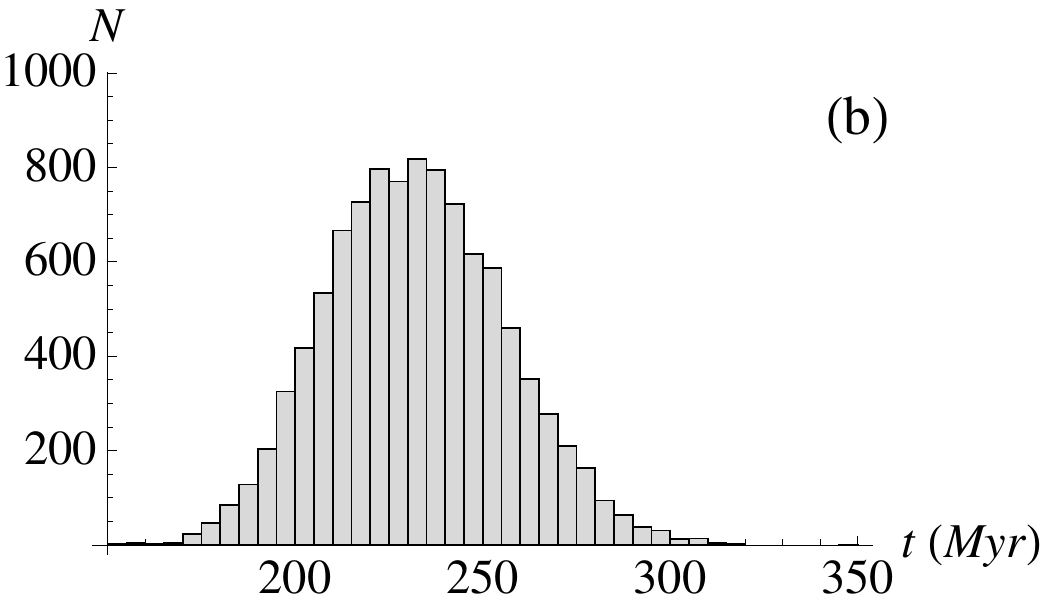}}
  \caption{(a) The distribution of decay times
  obtained from simulations for Class A particles (\S\ref{caseone}). Particles with
  $\rho = 1 \gcm$ and $r_g \approx 100 \mm$ were initialized with a semi-major axis of $3000 \AU$
  and an eccentricity uniformly selected on the interval (0,0.99). The
  simulation was run for $40 \Myr$, and a decay constant was obtained
  by fitting an exponential to the $a(t)$ dependence as shown in
  Figure \ref{decayfig}. (b) The distribution of decay
  times for particles with $\rho = 1 \gcm$ and $r_g \approx 0.1
  \text{ cm}$ (Class B parameters, see \S\ref{casetwo}). Eccentricities and
  semi-major axes are initialized
  in the same way as in (a), but the integration is now for 400 Myr and
  we have included the effects of passages through atomic clouds (see
  \S\ref{casetwo} for more details).} 
  \label{decayhist}
\end{figure}

The distribution of decay times $\gamma^{-1}$ is shown in 
Figure \ref{decayhist}b. The mean decay time for this simulation is 
$\langle t_\text{decay} \rangle = 232$ Myr with a standard deviation of $24$ Myr in agreement 
with analytical estimates. Note that the distributions of $t_\text{decay}$
for Class A and Class B particles look quite different. This is 
because for Class B there is an additional source of variance that is not
present for Class A: we model the passages through atomic clouds as 
stochastic events, and thus, although there are on average 8 cloud 
passages per simulation, the number and timing of passages
in any given simulation do not have to be the same. As a result,
the distribution of $t_\text{decay}$ for Class B particles is closer
to Gaussian.

We also give limits on $t_\text{decay}$ for Class B particles with
  starting semimajor axes at 300 AU and $10^4$ AU.
  For $\alpha = 0.05$ ($r_g = 20 \mm$) in an atomic
  cloud\footnote{The
    reason that particles with $\alpha = 0.05$ at $300$ AU in an atomic cloud
    (relevant for Class B) are almost
    same size as particles with $\alpha =0.05$ at $300$ AU in the warm phase
    (relevant for Class A) comes from the
    fact that the electric force starts to dominate the total drag in both
    environments for particles of size $\sim 10 \mm$, and we have
    assumed that particles are charged to $1$ Volt in every
    environment.} at $300$ AU the decay time is $ 2.8 \Myr <
t_{B,\text{decay}} < 5.2 \Myr$, so transport is extremely
  important for these particles if they are not destroyed on even
  smaller timescales (e.g. by collisions). On the other hand at $10^4$ AU and
  $\alpha = 0.05$ ($r_g = 1 \cm$), the decay time is $ 2.2 \Gyr <
t_{B,\text{decay}} < 2.6 \Gyr$ so transport is insignificant, since
  passages through molecular clouds happen on shorter timescales.

\subsection{Class C}
\label{casethree}

Class C particles survive passages through 
molecular clouds which implies that they have $T_s=4.5$ Gyr, i.e. the
age of the Solar System. At $3000 \AU$ a particle with $\rho = 1 \gcm$
and $r_g \approx 30 \cm$ has $\alpha = 0.05$ in a molecular cloud 
so we take these to be the parameters of a Class C particle. Using values for the filling 
fractions from Table \ref{ISMphases} and the formula
\ba
t_{C,\text{decay}} = 
\left(\displaystyle\sum_i \gamma_i f_i\right)^{-1},
\ea
where the sum runs
over the different phases of the ISM, and $\gamma_i$ and $f_i$ are
the corresponding orbital decay constants and filling factors, the
decay time is $ 33 \Gyr < t_\text{decay} < 56 \Gyr$, which is
considerably longer than $T_s$. A particle with $\alpha = 0.05$ in a
molecular cloud would need to have a semimajor axis of $1000$ AU for
the decay time to equal the age of the solar system (the size of
such a particle would be $r_g \approx 3 \cm$). Thus, 
beyond $1000$ AU, differential drag is 
unimportant for Class C particles, and they should remain at
essentially the same semi-major axis at which they were created.

\subsection{Comparison with previous work}
\label{subsubsect:comparison}

\citet{Sternerosion} has previously estimated the
perturbing force on particles in the Oort Cloud. However, he only
considered gas dynamical drag and neglected both the Coulomb drag and the
electric force. Moreover, his estimates for the gas drag
rely on the assumption that $v_w$ is much larger
than the thermal velocity of the species in the wind, i.e. $s \gg
1$. From Table \ref{ISMphases} we see that this assumption is
questionable for the warm phase and incorrect for the
coronal phase.

\citet{Sternerosion} has also considered the importance of a 
passing shock from a nearby
supernova for ejecting particles and has found that a supernova at 40 pc can
eject particles of radius $\sim 100 \mm$. Assuming a local supernova rate of
0.02 yr$^{-1}$, the average time between supernova explosions within 40 pc is
$\sim 40 \Myr$. This is similar to the timescale for passages through
atomic clouds, and because a typical class A particle has a radius of
$r_g \sim 100 \mm$, the effect of supernova explosions is simply to
reduce the average lifetime of the smaller Class A particles by about a factor
of two. 

Finally, \citet{Sternerosion} has pointed out that Oort Cloud objects
should be eroded by high velocity impacts with interstellar
grains. This is most relevant for Class A grains since they will be eroded within $40 \Myr$,
according to Stern's estimates. Thus it may be possible that these
particles are eroded to the point of ejection before their semi-major
axes can decay appreciably. However this statement
is speculative, because the erosion rate is not well understood and
depends on both the composition of the grains and their 3-dimensional structure.


\section{Application to Satellite Observations}
\label{satellites}

Dust experiments on board the {\it Ulysses} and {\it Galileo}
satellites have detected a significant flux of dust particles
which appear to be co-moving with the ISM flow through the Solar 
System: within the experimental uncertainties, which are quite 
significant\footnote{The accuracy of arrival direction determination is set by the 
opening angle of the dust detector, which is $140^\circ$ for 
Ulysses; particle speed can be determined with a $1\sigma$ 
accuracy of $\approx 7-8$ km s$^{-1}$ (Grun \etal 1994).}, 
dust grain velocities agree both in direction and magnitude with
${\bf v}_w$. 

A surprising feature of this dust component is that it contains 
a significant fraction of large grains with masses in excess of 
$> 3 \times 10^{-13}$ g, i.e. above the upper cutoff of the 
standard MRN dust size distribution (Mathis \etal 1977; 
Weingartner \& Draine 2001)  expected for the local ISM \citep{Landgraf}. 
\citet{Draineconundrum} has shown that if these massive particles 
are pervasive throughout the ISM, then (1) the mass locked
up in these large grains may be inconsistent with the amount 
of refractory material potentially available for dust grain formation 
in the Galaxy and, (2) even more importantly, the  
reddening curve calculated accounting for the large grain population 
is grossly inconsistent with observations. This essentially excludes 
the idea that large grains can be broadly spread through the Galaxy 
in the amounts measured by the satellites. At the same time, 
it has proven to be difficult to identify a mechanism that would 
cause a concentration of such large grains even locally in the
nearby ISM (Draine 2008).

Another possibility for the origin of massive grains is that they 
come from the OSS, e.g. get produced in cometary
collisions beyond the heliopause and become coupled to the ISM flow. 
This possibility has been first proposed by \citet{Frisch}, who 
dismissed it based on the supply argument: given the measured mass
flux in large grains and assuming a typical radius of the dust 
production region ($R_{\text{Oort}}=5\times 10^4$ AU 
in \citet{Frisch}) the whole 
mass of the Oort Cloud $M_{\text{Oort}} =20-40 \ M_{\earth}$ 
would be ground down in collisions on a timescale 
\ba
\label{oortlife}
t_{\text{Oort}} \sim \frac{M_{\text{Oort}}}{4 \pi R_{\text{Oort}}^2
  f_{\text{m}}}\lesssim 10^5\,\text{yr}, 
\ea
where $f_{\text{m}} = 2
\times 10^{-20} \text{ g} \text{ cm}^{-2} \text{ s}^{-1}$ is 
their assumed value for the flux of interstellar dust grains.
However it is unlikely that main dust production region in the
Oort Cloud lies at $5\times 10^4$ AU since the spatial distribution 
of comets is expected to be concentrated towards the Cloud's 
inner edge, which may lie at 3000 AU or possibly even closer 
\citep{Kaib,Kaib2,Dones}. If the inner edge is indeed at 
3000 AU, this increases the estimate of the 
Cloud lifetime to $t_{\text{Oort}} \sim 20 \Myr$, but still 
does not reconcile it with the age of the Solar System. For
the dust particles detected by satellites to have the Oort 
Cloud origin and for the Cloud to have a fragmentation timescale
$t_{\text{Oort}}$ on the order of several Gyr requires
the small dust particles (which get entrained in the ISM wind) 
to be mainly produced at distances of order $500$ AU from the BSS.

Our present study reveals two physical effects which may cause
the concentration of intermediate size grains (particles with sizes 
above the ejection threshold, which can produce dust grains with sizes
below the
ejection threshold in mutual collisions) towards the Inner Solar System.
The first effect is the secular variation of eccentricity in the Stark problem 
(see \S\S\ref{subsect:Stark_planar}-\ref{subsect:Stark_3D})
which in some cases leads to high values of $e$ and allows particles to
venture closer to the BSS. The second effect is the gradual 
decrease in the semi-major 
axes of grains caused by differential drag 
(see \S\ref{sect:orbdecay}), which again makes it possible for grains 
to come closer to the BSS. 

To investigate whether these effects can reconcile 
the observed flux of large ISM grains with their possible origin 
in the Oort Cloud it is necessary to nail down the location of
the inner edge, construct a model of collisional 
evolution of the Cloud, investigate the transport of ejected small 
particles towards the Inner Solar System, and so on. Given the complexity 
of associated modeling we do not pursue this effort here, leaving it
for future investigation. Such a study may potentially set useful
limits on the amount of dust material that is arriving into
the Inner Solar System from the Oort Cloud. Its results will also be
relevant for interpreting observations of interstellar meteoroids
(sizes below $100 \mm$) detected in radar observations 
\citep{Weryk,Murray}.


\section{Summary}
\label{sect:concl}

We have investigated the dynamics of dust particles with sizes between 
$\sim 1\,\mu$m and several meters in the Outer Solar System subject to the 
effects of the flow of interstellar gas beyond the heliopause. 
We analyzed various forces affecting the motion of small particles 
and showed that the electromagnetic force can be quite important for dust 
dynamics: the electric field induced in the Solar System frame by the 
magnetic field carried with the ISM flow can be the main determinant 
of the dynamics of dust particles which are bound to the Solar System. The magnetic 
force is important for motion of small particles which get ejected 
from the Solar System. These statements are true in particular 
for particles smaller than $100 \mm$
interacting with the warm phase of the ISM through which the Solar System is
currently passing. 
Drag forces -- Coulomb drag against the ionized component of the
ISM flow and gas drag -- are crucial for dust dynamics while
the Solar System is passing through molecular or atomic clouds. 
Radiation forces are never important for the dynamics of dust 
particles bigger than a micron in size outside of the heliosphere. 

We have demonstrated that the effect of these non-gravitational forces can be
well-described by the classical Stark problem, and have studied the
ejection conditions for dust grains using the approach based on the 
Kustaanheimo-Stiefel transformation. Based on these results, we
determined the particle sizes and semi-major axes at which they
are ejected from the Solar System in different ISM phases.
Rare passages through molecular clouds can eject particles as
large as 1 m if they were originally located at $10^4$ AU from the BSS.

We have also explored the motion of
bigger, bound grains using a perturbative approach based on
orbit-averaging the non-Keplerian part of the Hamiltonian. This 
allowed us to obtain a complete analytical description of the system
on timescales longer than an orbital period. We have shown, in particular, that the
eccentricity of dust particles oscillates in a regular fashion 
reaching high levels in some circumstances and allowing dust 
grains to explore radii much smaller than their semi-major axes. We have also demonstrated that
the component of the drag force which depends on particle velocity 
causes the decay of particle orbits, and this may be an important effect
for small grains.

Finally, we have discussed the possible relevance of these dynamical
effects for the origin of big grains recently discovered by 
the {\it Ulysses} and {\it Galileo} satellites, which appear to be 
flowing into the Solar System with the ISM gas.

\acknowledgements

We are indebted to Bruce Draine for instigating our interest
in this problem, continuous encouragement, numerous discussions, 
and advice. We thank Scott Tremaine and an anonymous referee for comments that helped improve the quality of the paper. This work has been financially supported by the 
Sloan Foundation and NASA grant NNX08AH87G.


\appendix


\section{Variations on the Stark Problem}
\label{starkvariations}

Our study until now has explicitly assumed that the Stark force is 
constant both in space and time, which in practice means that all 
our results are valid if ${\bf S}$ does not vary on timescales 
shorter than $\sim T_\text{stark}$. We now briefly discuss whether the 
particle semi-major axis can vary in a systematic fashion 
if we abandon the assumption of constant ${\bf S}$.  


\subsection{Radially-Dependent Grain Charging}
\label{eq:varcharge}

We have previously assumed that a particle's charge is fixed over 
its orbit. However, one may expect the charge to vary as a function of
distance from the Sun \citep{Kimura} because of increased exposure 
to ionizing radiation at smaller radii. Then even if the direction
and magnitude of the induced electric field are constant, the magnitude of 
the electric force and the Coulomb drag force will still vary as 
functions of distance from the Sun due to the radial dependence of 
the grain charge. 

\begin{figure}[h] 
  \centering
  \includegraphics[width=.5 \textwidth]{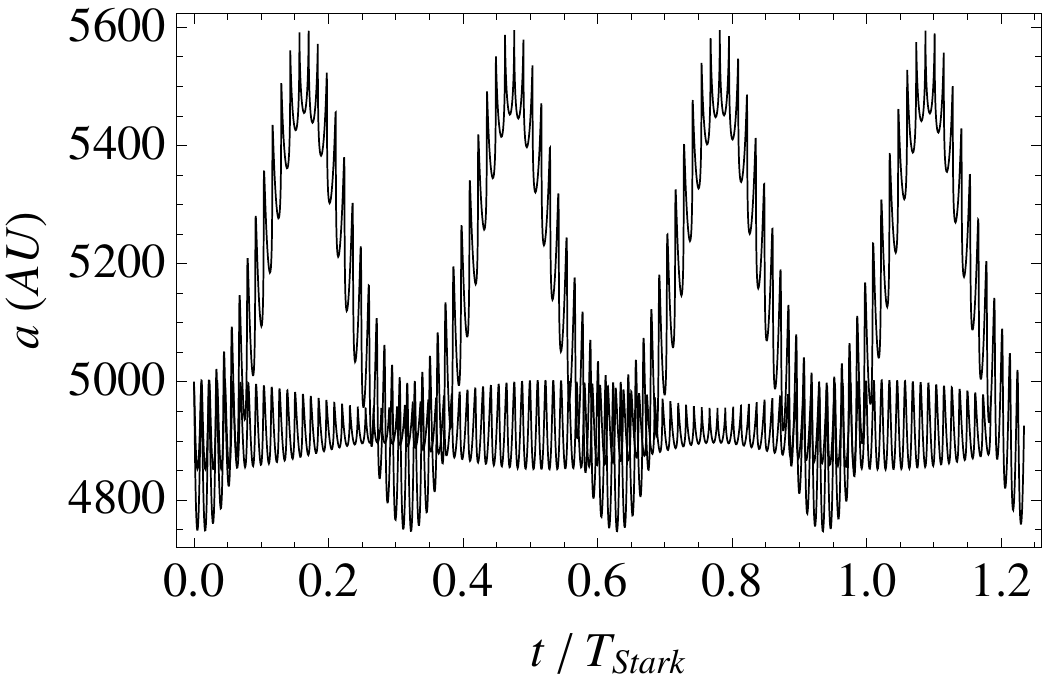}
  \caption{Evolution of the semi-major axis with radially-dependent
  grain charge $q(r) = q_0\left[1 + \ln (1+r_0/r)\right] $ (large
  amplitude curve) and with fixed grain charge $q_0$ and $\alpha =
  0.02$ (small amplitude
  curve). The initial semi-major axis of both grains was chosen to be
  $5000 \AU$, and the other orbital parameters were chosen randomly, but
  are the same for both simulations.}
  \label{graincharging}
\end{figure}

To investigate what effect a radially-dependent charge has on
the orbit, we allow the grain charge $q$ to vary according to a simple prescription 
\begin{equation}
\label{grainchargeq}
q(r) = q_0\left[1 + \ln (1+r_0/r)\right], 
\end{equation}
where $q_0$ and $r_0$ are constants.
Thus, the charge
tends to $q_0$ for $r \gg r_0$ and increases slowly as the grain
approaches
the Sun. We simulate the effect of grain charging numerically and
choose $r_0$ to be the starting
semi-major axis of the grain and $q_0$ such that the grain is
charged to 1 V at infinity. As shown in Figure \ref{graincharging} 
we find that
although grain charging does lead to a systematic change in the
semi-major axis on an orbital timescale, the changes are
periodic on a timescale of order $T_\text{stark}$ and on average
the semi-major axis stays constant. This is because radially-dependent grain
charging does no work on a particle over a closed Stark orbit and
cannot affect its semi-major axis in a systematic way.

\subsection{Time-Varying Stark Vector}
\label{subsubsect:stochStark}

If we allow the Stark vector to vary in magnitude and in time, it
will no longer be true that the semi-major axis or the energy of the orbit
are conserved. However, if the timescale on which ${\bf S}$ varies
is long compared to the orbital period $T_K$, then adiabatic 
invariance arguments suggest that the semi-major axis should not vary 
in a systematic fashion. Indeed, the classical Stark problem is 
separable in parabolic coordinates \citep{LL} and admits 
three adiabatic invariants (Banks \& Leopold 1978) which are conserved 
when ${\bf S}$ varies slowly. The energy of a particle can be expressed
as a function of these invariants and the value of ${\bf S}$, meaning 
that as the ambient magnetic field and wind velocity slowly change their
magnitude and direction around some mean values, the semi-major axis
will simply oscillate around its mean value as well.

At the opposite extreme,
${\bf S}$ may have a stochastic component which varies rapidly on
timescales shorter than $T_K$, e.g. due to the small scale 
magnetic turbulence or random charge fluctuations on the grain.
In this case we expect a random walk-like evolution of the semi-major 
axis to take place even if the stochastic component of 
${\bf S}$ averages to zero. A similar effect of a random walk in 
the semi-major axes of comets caused by random stellar passages through 
the OSS has been previously explored in Heisler \& 
Tremaine (1986). The resulting diffusion of 
the initial distribution of particle semi-major axes
will bring some particles to smaller $a$ and make them more 
gravitationally bound; others will move to higher $a$, 
and possibly become unbound. We expect the smallest bound dust grains with 
large initial semi-major axes to be most affected by diffusive
evolution from a rapidly fluctuating Stark vector. 

We do not attempt to model the random walk evolution here because 
of the large uncertainties in the input ingredients that such a 
model will involve: spectrum and strength of magnetic turbulence,
random fluctuations of the grain charge, etc. We leave this
subject for future investigation.  We do point out, though, that
  diffusion caused by a rapidly fluctuating Stark vector would add on
  top of diffusion caused by random stellar perturbations. 
\citet{Kaib2} have performed simulations which include the effects of stellar
  perturbations and have found that diffusion is ineffecient
  inside of $10^4$ AU; only $10-15 \%$ of objects having $a >
  2\times 10^4$ AU arrived there by diffusion from initial orbits 
having $a < 10^4$ AU after a time of 4.5 Gyr. Thus, diffusion caused 
by stellar perturbations alone
  will not cause significant transport in semimajor axis at the inner edge of the Oort
  Cloud, where most of the dust is likely to be concentrated. 


\section{Ejection Criterion}
\label{app:ejection} 

To understand the ejection conditions in the Stark problem, we use the
analytical
results of Kirchgraber (1971) and Dankowicz (1994) which rely on the
use of KS transformation proposed by Kustaanheimo \& Stiefel (1965).
This transformation relates six components of the 3-dimensional Cartesian 
coordinates ${\bf x}$ and velocities ${\bf v}$ to 8 components of 4-dimensional generalized 
coordinate vector ${\bf u}$ and velocity ${\bf w}$ via a non-linear 
prescription, see e.g. Kirchgraber (1971). Upon transformation to KS
variables, the standard Keplerian
problem reduces to the problem of harmonic oscillator motion in
four dimensions, which is easier to treat in some cases.
  
In the case of the Stark problem, one can show that determining whether 
a given particle is bound or not to a gravitating center is 
equivalent to studying the problem (Kirchgraber 1971; Dankowicz 1994)
of motion in $R$-space ($R \equiv u_1^2+u_4^2=(r+x_1)/2$, $u_1$ and 
$u_4$ are the first and the fourth components of generalized
coordinate vector ${\bf u}$) 
with zero energy 
\ba
\left(R^\prime\right)^2+V(R)=0
\label{eq:genpot}
\ea
in the potential (Kirchgraber 1971)
\begin{equation}
V(R)=-2SR^3+4hR^2-8 \tilde E R+4j^2,
\label{eq:pot}
\end{equation}
where, based on the initial conditions for ${\bf x}^0=(x_1^0,x_2^0,x_3^0)$
and ${\bf v}^0=(v_1^0,v_2^0,v_3^0)$ (axis 1 is in the direction of the 
Stark vector which has components ${\bf S}=(S,0,0)$) we have the 
following expression for various constant factors entering this 
potential:
\begin{eqnarray} 
&& j=\frac{1}{4}\left(v_2^0 x_3^0-v_3^0 x_2^0\right)=-\frac{J_z}{4},
\label{eq:j}\\
&& \tilde E=\frac{1}{8}\left[\frac{v_1^{0 2}(r^0+x_1^0)}{2}+
\frac{(v_2^{0 2}+v_3^{0 2})(x_2^{0 2}+x_3^{0 2})}
{2(r^0+x_1^0)}+v_1^0(x_2^0v_2^0+x_3^0v_3^0)\right]\nonumber\\
&& +\frac{h}{4}(r^0+x_1^0)-\frac{S}{16}(r^0+x_1^0)^2,
\label{eq:E}\\
&& h=-\frac{1}{2}\left[\frac{v^{0 2}}{2}-
\frac{GM}{r^0}+V_\text{stark}\right],~~~~~
V_\text{stark}=-Sx_1^0\label{eq:h}.
\end{eqnarray}
Constants $j$ and $h$ are integrals of motion arising from the 
conservation of the $z$-component of the angular momentum 
and energy. The constant $\tilde E$ is also an integral of motion which is 
generic for Stark problem \citep{LL}. It can be rewritten
as
\ba
\tilde E=\frac{1}{8}\left[GM-e_z-\frac{S\rho^2}{2}\right],
\ea
where $\rho=\sqrt{x_2^2+x_3^2}$ is the component of ${\bf r}$ perpendicular to $z$-axis
(or axis 1), and $e_z$ is the $z$-component of the eccentricity vector
${\bf e}\equiv {\bf v}\times\left({\bf r}\times 
{\bf v}\right)-GM{\bf r}/r$ -- an integral of motion unique to the pure
Keplerian problem.

One can show that whenever the physical distance $r$ becomes infinite
(the particle is unbound), the coordinate $R$ also becomes infinite (the 
opposite is also true -- finite $R$ corresponds to finite $r$), which 
makes finiteness of $R$ an indicator of the boundedness of 
particle. For this reason we only need to determine under which conditions there 
exist bound states for zero-energy motion in the potential $V(R)$ ---
their existence and particle location in one of these states would 
be equivalent to it being bound.

For the problem of motion in a cubic potential represented by equations
(\ref{eq:genpot})-(\ref{eq:pot}) at $R>0$ 
(only positive $R$ is physically interesting) bound states appear whenever  
$V(R)$ has 3 roots all of which are positive -- then there is a 
valley in the potential curve in which the motion is bound. 
This is possible if the 
two extrema of $V(R)$, a minimum and a maximum occur at non-negative $R_-$
and $R_+$, and $V(R_-)\le 0$ while $V(R_+)>0$. It is easy to see that
\begin{equation}
R_\pm=\frac{2}{3}\frac{h}{S}\left[1\pm\sqrt{1-3\frac{\tilde ES}{h^2}}\right].
\label{eq:Rpm}
\end{equation}

To be in the valley, the particle also needs to have $R^0<R_+$.
Thus, a particle is {\it bound} if and only if for given ${\bf x}^0,{\bf v}^0$ we have
\begin{equation}
R_-\ge 0,~~~R_+>0,~~~V(R_-)\le 0,~~~V(R_+)>0, ~~~\mbox{and}~~~ 
R^0=\frac{r^0+x_1^0}{2} < R_+.
\label{eq:bound}
\end{equation}
In practice, for given ${\bf x}^0,{\bf v}^0$ one 
needs to (1) calculate the constants $j,\tilde E,h$ using equations 
(\ref{eq:j})-(\ref{eq:h}), (2) calculate $R_-$ and $R_+$ using equation 
(\ref{eq:Rpm}), and (3) check whether all of the conditions 
(\ref{eq:bound}) are satisfied. If they are, then the motion is
bound, but if at least one of them is violated then the motion 
is {\it unbound}.

We can directly use equations (\ref{eq:pot}) - (\ref{eq:h}) 
to find the ejection criterion
for particles with zero initial velocity, $v_1^0 = v_2^0 =
v_3^0 = 0$. In this case, the constants of the motion are $j = 0, ~ \tilde E
= -(r^0+x^0)[Sr^0(r^0-x^0)-2GM]/16r^0, ~ h = (Sx_1^0+GM/r^0)/2$. Setting
$V(R)=0$ to find the turning points of the motion and solving the
resulting quadratic equation for $R$, the condition for ejection 
becomes $(r^{0})^2 >GM/S$. Since the initial semi-major axis $a^0 = r^0/2$ in 
the zero velocity case, ejection must occur for $\alpha> 0.25$,
as given by equation (\ref{eq:ejsimple}).    


\section{Numerical Integrator}
\label{WHintegrator}

To assist our study of orbital motion, we have written an integrator
based on Burdett-Heggie (BH) regularization. We start from the BH regularized equation of motion
\citep{galacticdynamics}
\begin{equation}
\label{regularized}
\frac{d^2 {\bf r}}{ds^2} = 2E {\bf r} - {\bf e} + r^2 \frac{{\bf F}}{m}, 
\end{equation}
where $ds = dt/r$, $\bf{e}$ is the eccentricity vector, $E$ is the
energy of the orbit, and ${\bf F}$ is the perturbing force, which need
not be small. We implement equation (\ref{regularized})
in a manner
similar to leapfrog \citep{galacticdynamics}, so our algorithm consists of a series of
${\bf D}_{1/2} {\bf K} {\bf D}_{1/2}$ steps, where ${\bf K}$ denotes a
full kick step and ${\bf D}_{1/2}$ denotes a half drift step. Each drift step
corresponds to analytically integrating equation (\ref{regularized})
without the perturbing forces for
a duration of regularized time $\Delta s / 2$, and
thus is simply the solution to an initial value problem for a forced
harmonic oscillator. Each kick step is then simply an instantaneous
application of the perturbing forces.

Our algorithm bears a similarity to the Wisdom-Holman
integrator \citep{Wisdom}, and in particular to the KS regularized
Wisdom-Holman integrator of \citet{Mikkola}. \citet{Preto} have also
devised a class of algorithms with an adaptive timestep for separable
Hamiltonians, and the integrator in this class which has a timestep
proportional to $1/r$, the same as regularization, follows a Keplerian
orbit exactly except for errors in the phase. The reason for
developing a new algorithm is so that we can easily treat the magnetic
force on particles.
Although we have found that the magnetic force is dynamically
unimportant for bound particles, because it is much smaller than the
electric force, it is important for ejected particles, and a future
study of the trajectories of these particles would need to include
this force. This rules out the algorithm of
\citet{Preto}, since the Hamiltonian is no longer separable in the
presence of a magnetic field. On the other hand, it should be possible
to modify the algorithm of \citet{Mikkola} to include magnetic fields,
but because both time and space coordinates are changed in KS
regularization, whereas only the time coordinate is changed in BH
regularization, it is simpler to treat magnetic fields using BH
rather than KS regularization. For this reason we have
decided to write our own algorithm rather than
modifying the one of \citet{Mikkola}.

One possible second order discretization for the update of the
(unregularized) velocity during the kick step is
\begin{equation}
  \label{vadvance}
  \frac{{\bf v}_{+\epsilon} - {\bf v}_{-\epsilon}}{\Delta t} =
  \frac{q}{m} \left( {\bf E}+\frac{{\bf v}_{+\epsilon} + 
  {\bf v}_{-\epsilon}}{2c} \times {\bf B} \right)
  + \frac{1}{m}{\bf F}_\text{drag} \left(\frac{{\bf v}_{+\epsilon} + {\bf
  v}_{-\epsilon}}{2} + { \bf v}_w \right),
\end{equation}
where ${\bf v}_{-\epsilon}$ and ${\bf v}_{+\epsilon}$ are the
velocities right before and right after the kick step, and the
magnitude of ${\bf F}_\text{drag}$ is given by equation (its direction is
of course opposite to the sum of orbital and wind velocities). Because
the kick step is assumed
to happen instantaneously, $r$ is constant during the kick step, and
we can convert between the unregularized velocity $dr/dt$, and the
regularized velocity $dr/ds$ using the simple transformation $ds =
dt/r$. 

One complication with implementing equation (\ref{vadvance}) is that it is implicit since
${\bf v}_{+\epsilon}$ appears on the right hand side. However, in the case
of a pure electromagnetic force with no
drag, equation (\ref{vadvance}) can be inverted to obtain an explicit scheme for the
particle advance. This is known as Boris's algorithm and is widely
used in particle in cell simulations of plasmas \citep{Birdsall}. The
kick step of Boris's algorithm can be represented as $ {\bf
  v}_{+\epsilon} = {\bf E}_{1/2} {\bf B} {\bf E}_{1/2}
  {\bf v}_{-\epsilon}$, where ${\bf E}_{1/2}$ represents half of an electric
acceleration step, and ${\bf B}$ is a full magnetic rotation step. This
algorithm is explicit, second order, time-reversible, and conserves energy.

Rather than solving the implicit equation (\ref{vadvance}), we extend Boris's
  algorithm to include the drag force by writing ${\bf v}_{+\epsilon} =
{\bf F}_{\text{D},1/2} {\bf E}_{1/2} {\bf B} {\bf E}_{1/2} {\bf F}_{\text{D},1/2}
  {\bf v}_{-\epsilon}$, where $ {\bf F}_{\text{D},1/2}$ is half of a drag
acceleration step. Such an algorithm is entirely explicit, and in the
  case where the drag force does not depend on the particle velocity, the expression
for the kick step above is identical to Boris's algorithm with the
substitution $q{\bf E} \to q{\bf E} + {\bf F}_\text{drag}(v_w)$. This is
not true in the general case for which the drag force can depend on
the orbital velocity, so we must test our algorithm to make sure that
it gives reliable results. 

The first test we perform is the planar Stark problem, so there is a
constant perturbing force in the plane of the orbit. The performance
of a variety of modified Wisdom-Holman integrators on this problem has
been studied by \citet{Rauch}. They found that for the energy
error to be bounded, the pericenter passage must be resolved. Since
the pericenter passage lasts for a time $t_p \propto r_p^{3/2}$,
where $r_p$ is the radius at pericenter, regularized integrators
with a timestep $\Delta t \propto r^{-1}$,
performed better than unregularized ones, when the eccentricity
approached unity.

\begin{figure}[h] 
  \centering
  \subfigure{\includegraphics[width=.55\textwidth]{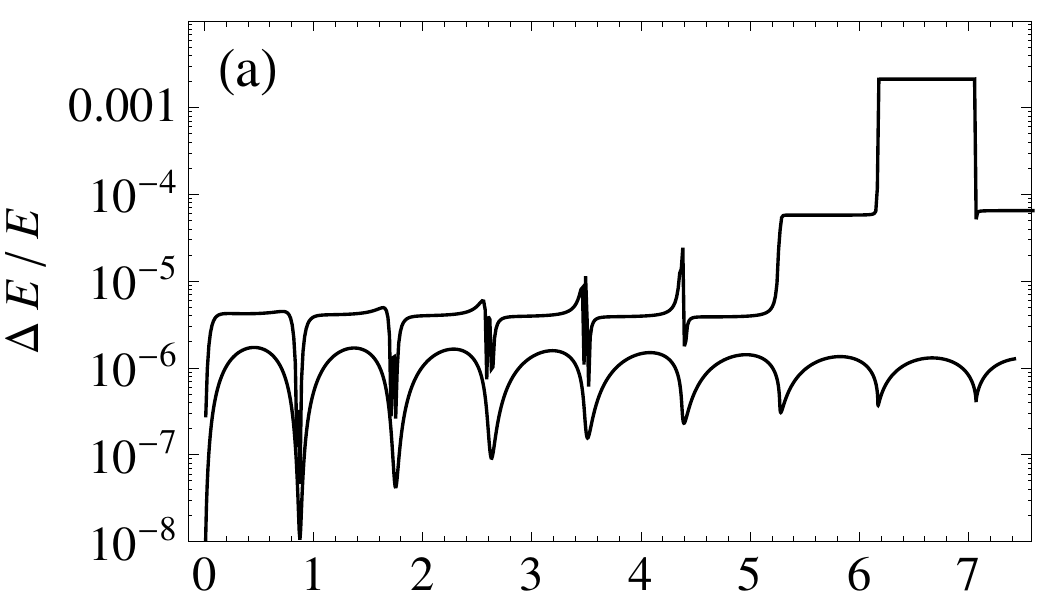}}
  \subfigure{\includegraphics[width=.55\textwidth]{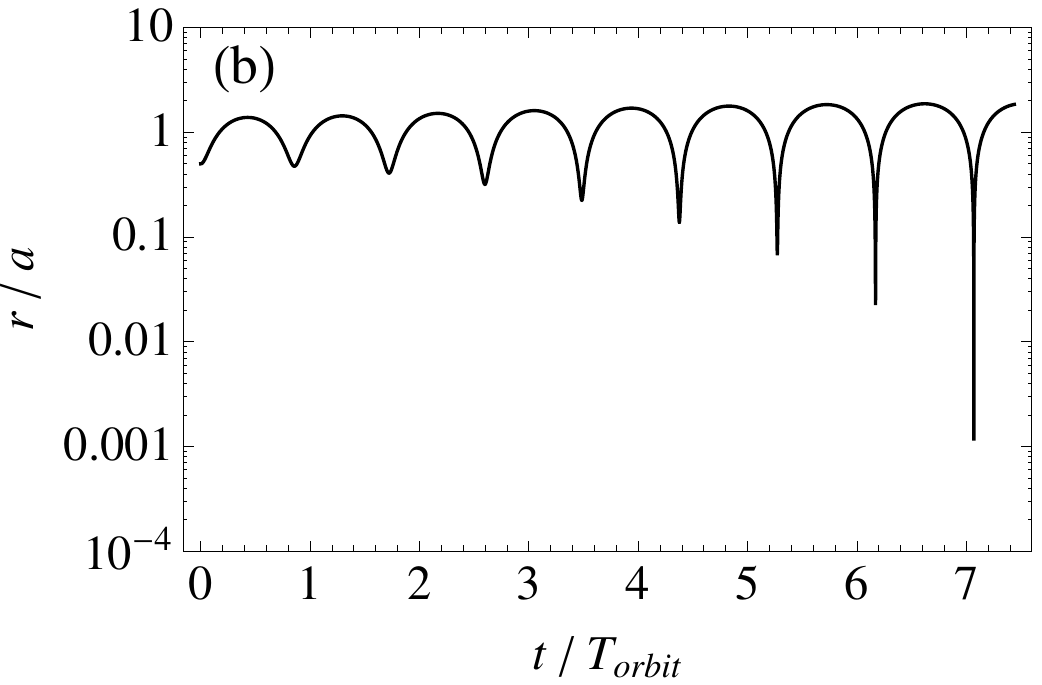}}
  \caption{(a) The fractional energy
  error of our BH-regularized integrator (lower curve)
  and an unregularized Wisdom-Holman integrator (upper curve) for the
  planar Stark problem with an initial
  eccentricity of 0.5, and $\alpha = 0.02$. Both simulations were run for a total of $\sim 2000$
  steps. (b) The ratio of the radius to
  the semi-major axis as a function of time for the same
  simulation shows that when the eccentricity becomes large and the particle
  passes very close to the force center, the error in the
  unregularized integrator fails to stay bounded, whereas it does stay
  bounded for our integrator.}
  \label{energycomparison}
\end{figure}

\begin{figure}[h] 
  \centering
  \subfigure{\includegraphics[width=.4\textwidth]{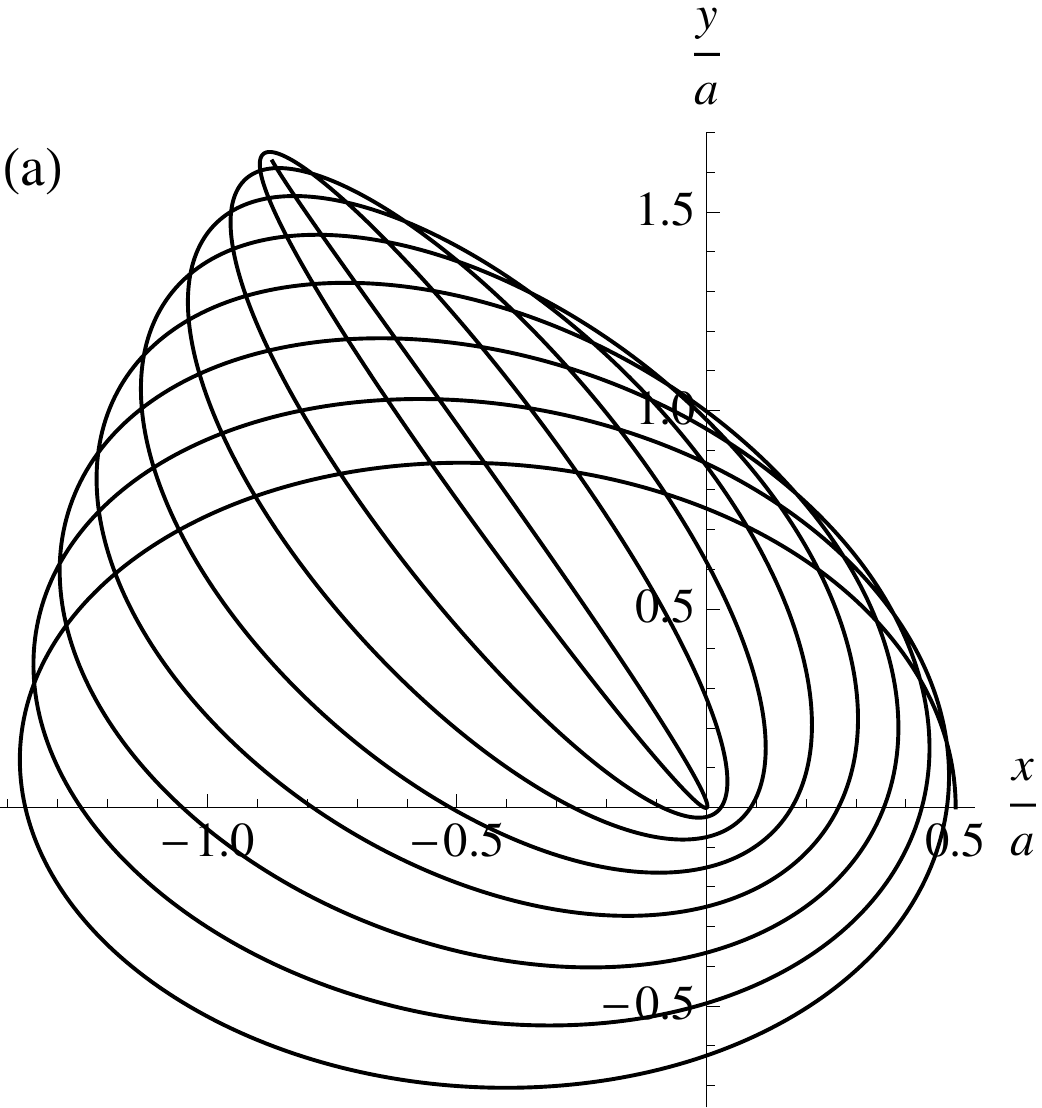}}
  \subfigure{\includegraphics[width=.4\textwidth]{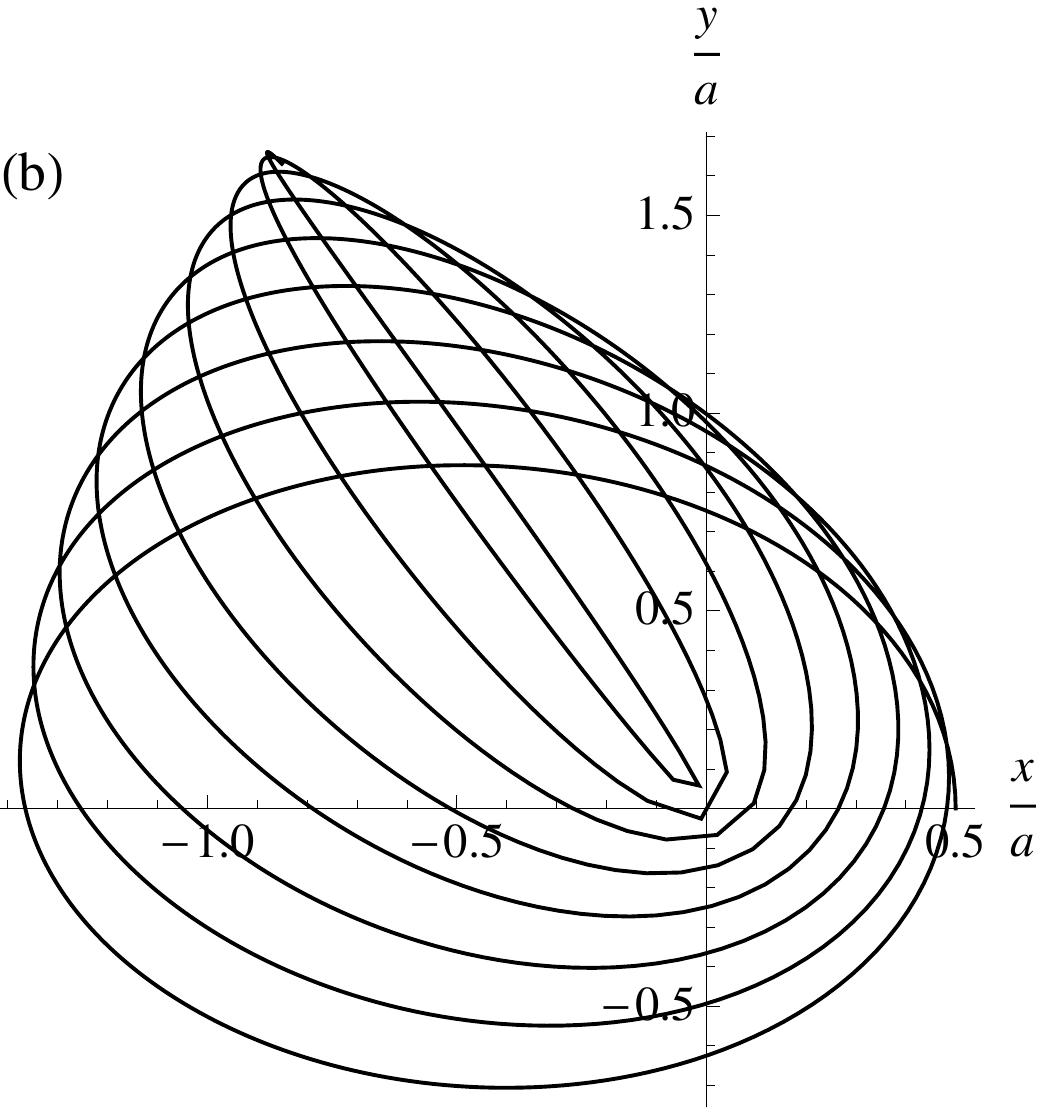}}
  \caption{Trajectory of a planar Stark orbit
  which has the perturbing force directed along the x-axis, an initial
  eccentricity of 0.5, and $\alpha = 0.02$, for two different
  integrators: (a) was computed using our
  BH-regularized integrator, and (b) was computed using an
  unregularized Wisdom-Holman integrator. Both orbits were computed
  using $\sim 2000$ timesteps for the whole simulation (not per orbit), but our
  integrator has much better resolution near periapse when the
  eccentricity is high.}
  \label{orbitcomparison}
\end{figure}

We compare
the performance of our algorithm with that of an unregularized
Wisdom-Holman integrator and find that it is much better at conserving
energy for highly eccentric orbits (Figure
\ref{energycomparison}). The reason for this is the much
finer resolution of our algorithm when passing close to the force
center, which can be seen in
Figure \ref{orbitcomparison}. 

We also test
that our algorithm can capture the effects of differential drag
by making sure that the solutions it gives are accurate and converged.
To test convergence, we study the decay rate of a Class A particle
(\S\ref{caseone}) with varying resolution. We find that the change in the
decay constant from 50 to 1000 timesteps per orbit is less than a
tenth of a percent. Because we typically use around 1000
timesteps per orbit, we can confidently say that we are able
to resolve the decay due to differential drag. 

\begin{figure}[h] 
  \centering
 \includegraphics[width=.5 \textwidth]{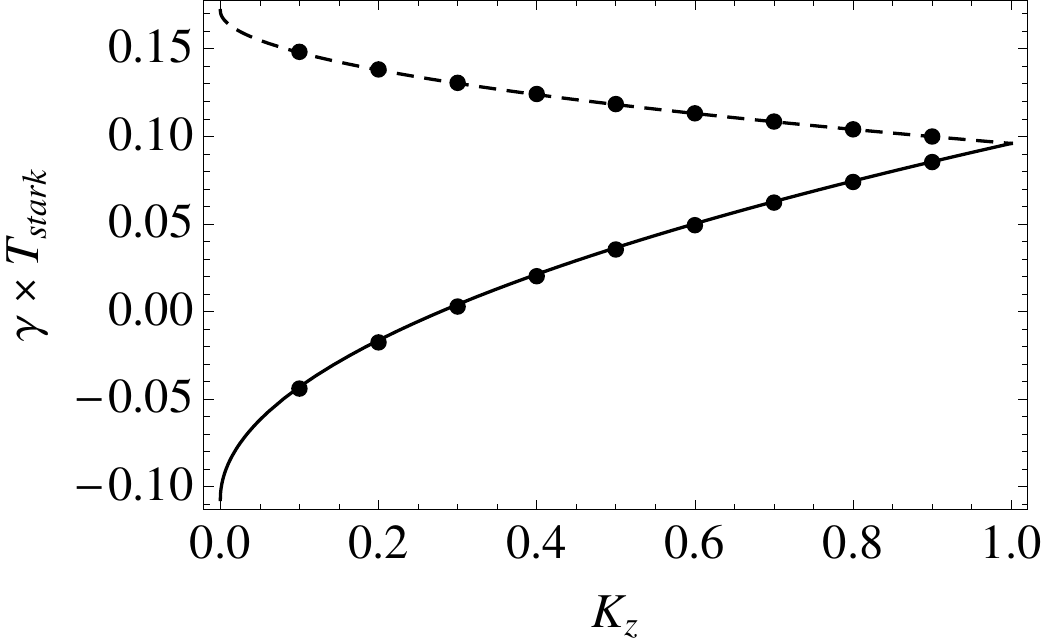}
  \caption{Plot of the dimensionless decay parameter $\gamma
  T_\text{stark}$ as a function of $K_z$ for the stationary orbit. The
  curves are analytical estimates for pure gas drag (dashed) and
  pure Coulomb drag (solid) obtained using equations (\ref{axisdecay}),
  (\ref{decayconstant}), and (\ref{stationarybeta}). The solid points are
  results obtained from simulations and agree with analytical
  estimates to within about a percent or less.} 
  \label{stationarydecayfig}
\end{figure}

To test that the code gives not
only a converged, but also an accurate result for the decay, we
compare its results against analytical estimates for stationary orbits
\S\ref{subsect:Stark_3D}. 
In that case we can precisely evaluate the factor $\gamma$ in equation
(\ref{decayconstant}) as a function of $K_z$. Orienting the Stark vector
along the z-axis we have the relations $({\bf v}_w \cdot
  {\bf \hat{a}}/v_w)^2 = (\sin i \sin \omega)^2$ and $({\bf v}_w \cdot
  {\bf \hat{b}}/v_w)^2 = (\sin i \cos \omega)^2$, which we can
use together with equation (\ref{eq:C^2}) and $K = \sqrt{K_z}$ to obtain 
\begin{equation}
\label{stationarybeta}
\beta = 1-\sqrt{K_z}
\end{equation}
From this relation for $\beta$ and definition (\ref{decayconstant}), 
it is straightforward to obtain
$\gamma T_\text{stark}$ and compare this with $\gamma T_\text{stark}$
from simulations integrated for one Stark period for pure gas drag and
pure Coulomb drag. The results are
plotted in Figure \ref{stationarydecayfig} and show that the simulations 
give not only a converged but also an accurate result for the
differential drag.

\begin{thebibliography}{}

\bibitem[Antonov 
\& Latyshev(1972)]{Antonov} Antonov, V.~A., \& Latyshev, I.~N.\ 1972, The Motion, Evolution of Orbits, and Origin of Comets, 45, 341 


\bibitem[Baines et al.(1965)]{Baines} 
  Baines, M.~J., Williams, 
  I.~P., \& Asebiomo, A.~S.\ 1965, \mnras, 130, 63 

\bibitem[Bahcall(1984)]{Bahcall} Bahcall, J.~N.\ 1984, \apj, 
  276, 169 

\bibitem[Banks \& Leopold(1978)]{Banks} 
  Banks, D., \& Leopold, J.~G.\
  1978 J. Phys. B., 11, 37

\bibitem[Binney \& Tremaine(2008)]{galacticdynamics} 
  Binney, J., \&
  Tremaine, S.\ 2008, {\it Galactic Dynamics}; Princeton Univ. Press

\bibitem[Birdsall \& Langdon(2005)]{Birdsall}
  Birdsall, C.~K., \& Langdon, A.~B.\ 2005, {\it Plasma Physics via
  Computer Simulation}; Taylor \& Francis Group

\bibitem[Brasser et al.(2006)]{Brasser} Brasser, R., Duncan, 
  M.~J., \& Levison, H.~F.\ 2006, Icarus, 184, 59 

\bibitem[Burns et al.(1979)]{Burns} 
  Burns, J.~A., Lamy, 
  P.~L., \& Soter, S.\ 1979, Icarus, 40, 1

\bibitem[Dankowicz(1994)]{Dankowicz} 
  Dankowicz, H.\ 1994, 
  Celestial Mechanics and Dynamical Astronomy, 58, 353 

\bibitem[Dones et al.(2004)]{Dones} Dones, L., Weissman, 
  P.~R., Levison, H.~F., 
  \& Duncan, M.~J.\ 2004, Star Formation in the Interstellar Medium: In Honor of David Hollenbach, 323, 371 

\bibitem[Draine \& Salpeter(1979)]{Drainedrag} 
  Draine, B.~T., \& Salpeter,
  E.~E.\ 1979, \apj, 231, 77

\bibitem[Draine(2009a)]{Draineconundrum} 
  Draine, B.~T.\ 2009, Space 
  Science Reviews, 143, 333  

\bibitem[Draine(2010)]{Drainebook} 
  Draine, B.~T. 2010, {\it Physics of
  the Interstellar and Intergalactic Medium}; preprint

\bibitem[Fernandez(1999)]{Fernandez} 
  Fernandez, J. A. 1999, in {\it Encyclopedia of the Solar System}, Academic Press; 537

\bibitem[Frisch et al.(1999)]{Frisch} 
  Frisch, P.~C., et al.  1999, \apj, 525, 492

\bibitem[Frisch et al.(2009)]{Frisch2} 
  Frisch, P.~C., et al. 2009, Space Sci. Rev., 146, 235

\bibitem[Heisler \& Tremaine(1986)]{Heisler} 
  Heisler, J.,
  \& Tremaine, S.\ 1986, Icarus, 65, 13  

\bibitem[Horanyi (1996)]{Hor1} 
  Horanyi, M.\ 1996, ARA\&A, 34, 383 

\bibitem[Horanyi et al.(1992)]{Hor2} 
  Horanyi, M., Burns, J. A., \& Hamilton, D. P.\ 1992, Icarus, 97, 248 

\bibitem[Kaib 
\& Quinn(2008)]{Kaib2} Kaib, N.~A., \& Quinn, T.\ 2008, Icarus, 197, 221 

\bibitem[Kaib \& Quinn(2009)]{Kaib} 
  Kaib, N.~A., \& Quinn, T.\ 2009,
  Science, 325, 1234 

\bibitem[Kimura \& Mann(1998)]{Kimura} 
  Kimura, H., \& Mann, I.\ 1998,
  \apj, 499, 454

\bibitem[Kirchgraber(1971)]{Kirchgraber} 
  Kirchgraber, U.~R.~S.\ 
  1971, Celestial Mechanics, 4, 340 
  
\bibitem[Kr{\"u}ger \& Gr{\"u}n(2009)]{Kruger} 
  Kruger, H. \& Grun, E.\ 2009, Space Sci. Rev., 143, 347

\bibitem[Kustaanheimo \& Stiefel(1965)]{KS} 
  Kustaanheimo, P. \& Stiefel, E.\ 1965, J. Reine Angew. Math., 218 

\bibitem[Landau \& Lifshitz(1976)]{LL} Landau, L.~D. \& Lifshitz,
  E.~M.\ 1976, {\it Mechanics}; Pergamon Press 

\bibitem[Landgraf(2000)]{Landgraf1} 
  Landgraf, M. \ 2000, \jgr, 105, 10303 

\bibitem[Landgraf et al.(2000)]{Landgraf} 
  Landgraf, M.,   Baggaley, W.~J., Gr{\"u}n, E., Kr{\"u}ger, H., 
\& Linkert, G.\ 2000, \jgr, 105, 10343

\bibitem[Mathis et al.(1977)]{Mathis} Mathis, J.~S., Rumpl, 
W., \& Nordsieck, K.~H.\ 1977, \apj, 217, 425

\bibitem[Mignard 
\& Henon(1984)]{Mignard} Mignard, F., \& Henon, M.\ 1984, Celestial Mechanics, 33, 239 

\bibitem[Mikkola(1997)]{Mikkola} Mikkola, S.\ 1997, Celestial 
Mechanics and Dynamical Astronomy, 67, 145 

\bibitem[Moro-Mart{\'{\i}}n 
\& Malhotra(2002)]{MoroMartin} Moro-Mart{\'{\i}}n, A., \& Malhotra, R.\ 2002, \aj, 124, 2305 

\bibitem[Murray \& Dermott(2001)]{MD} 
Murray, C.~D. \& Dermott, S.~F. 2001, {\it Solar System Dynamics}; 
Cambridge University Press

\bibitem[Murray et al.(2004)]{Murray} Murray, N., Weingartner, 
J.~C., \& Capobianco, C.\ 2004, \apj, 600, 804 

\bibitem[Opher et al.(2009)]{Opher} 
  Opher, M., Alouani Bibi, F., Toth, G., Richardson, J. D., Izmodenov,
  V. V., \& Gombosi, T. I. 2009, Nature, 462, 1036

\bibitem[Pastor et al.(2010)]{Pastor} Pastor, P., Klacka, J., 
\& Komar, L.\ 2010, arXiv:1008.2484

\bibitem[Preto 
\& Tremaine(1999)]{Preto} Preto, M., \& Tremaine, S.\ 1999, \aj, 118, 2532

\bibitem[Rauch \& Holman(1999)]{Rauch} 
  Rauch, K.~P., \& Holman, M.\ 1999, \aj, 117, 1087 

\bibitem[Richardson(2009)]{Richardson} 
Richardson, J. D. \& Stone, E. C. 2009, Space Sci. Rev., 143, 7 

\bibitem[Scherer(2000)]{Scherer} Scherer, K.\ 2000, \jgr, 105, 
10329 

\bibitem[Stark \& Kuchner(2009)]{Stark} 
Stark, C. C. \& Kuchner, M. J. 2009, ApJ, 707, 543

\bibitem[Stern(1988)]{Sterncollisions} 
  Stern, S.~A.\ 1988, Icarus, 73, 
  499 

\bibitem[Stern(1990)]{Sternerosion} 
  Stern, S.~A.\ 1990, Icarus, 84, 447

\bibitem[Weingartner \& Draine(2001)]{Weingartner} 
  Weingartner, J.~C.,
  \& Draine, B.~T.\ 2001, \apj, 553, 581

\bibitem[Weryk \& Brown(2004)]{Weryk} 
  Weryk, R.~J., \& Brown, P.\ 2004, Earth Moon and Planets, 95, 221 

\bibitem[Wisdom \& Holman(1991)]{Wisdom} 
  Wisdom, J., \& Holman, M.\ 1991, \aj, 102, 1528
	
\bibitem[Wyatt(2008)]{Wyatt} 
  Wyatt, M. C. 2008, ARA\&A, 46, 339

\end{thebibliography}
\end{document}